\documentclass[preprint,12pt]{elsarticle}

\usepackage{amssymb}
\usepackage{comment}
\usepackage{hyperref}
\usepackage{floatrow}
\usepackage{algorithm}
\usepackage{graphicx}
\usepackage{algpseudocode}
\usepackage{amsmath}
\usepackage[size=footnotesize]{subcaption}
\graphicspath{{figures/}}
\usepackage{cleveref}
\crefname{appendix}{}{}
\usepackage{soul}
\usepackage{geometry}
\usepackage{algorithmicx}
\usepackage[dvipsnames]{xcolor}
\usepackage{booktabs}
\usepackage{longtable}
\usepackage{siunitx}
\usepackage{stmaryrd}
\usepackage{algpseudocode}
\usepackage{algorithm}
\usepackage{adjustbox}
\usepackage{color, colortbl}
\usepackage{multirow}
\usepackage{xcolor}
\usepackage{ulem}
\usepackage{pgfgantt}
\usepackage{microtype}
\usepackage{tabularx}
\usepackage{subcaption}
\usepackage{import}     
\usepackage{transparent}

\newcommand{\x}{\mathbf{x}}

\usepackage{floatrow}
\usepackage{siunitx}
\journal{Engineering Applications of Computational Fluid Mechanics}

\begin{document}

\begin{frontmatter}


\title{Evaluating simulation techniques for lubricant distribution in gearboxes}

\author[inst1,inst2]{Pawan S. Murthy}
\affiliation[inst1]{organization={Robert Bosch GmbH, Corporate Research},
            addressline={Robert-Bosch-Campus 1}, 
            city={Malmsheim},
            postcode={71272}, 
            country={Germany}}
\author[inst1]{Anja Lippert \corref{cor1}}

\author[inst2]{Andrea Beck}
\affiliation[inst2]{organization={University of Stuttgart, Institute of Aerodynamics and Gas Dynamics},
            addressline={Wankelstrasse 3}, 
            city={Stuttgart},
            postcode={70563}, 
            country={Germany}}

\cortext[cor1]{Corresponding author: Anja.Lippert@de.bosch.com}

\begin{abstract}

Efficient lubrication is crucial for the performance and durability of high-speed gearboxes, particularly under varying load conditions. Excess lubrication leads to increased churning losses, while insufficient lubrication accelerates wear on contact surfaces. Due to the high rotational speeds involved, direct experimental visualization of lubricant distribution within gearboxes is challenging, making numerical simulations indispensable. Although various modelling approaches exist, a direct comparison that jointly evaluates accuracy and computational efficiency is missing. Furthermore, studies on the computational modelling of highly viscous lubricants such as grease in gearboxes are limited.

This study addresses these gaps by comparing two mesh-based Eulerian solvers (OpenFOAM and Ansys-Fluent) and two Lagrangian particle-based solvers (PreonLab and MESHFREE) for oil distribution in gearboxes. Two benchmark cases are considered: one for qualitative assessment and another for quantitative evaluation. OpenFOAM and Ansys-Fluent show good agreement with the experiment data in selected cases, but incur a significant computational cost. PreonLab performs well qualitatively, yet exhibits greater deviation in quantitative predictions. These comparisons provide information for selecting the suitable solver according to specific simulation requirements. Furthermore, the study extends to grease distribution by first validating the solver and then investigating the influence of filling volume and gear speed on the amount of grease deposited on gears. The benchmark cases presented provide a reference framework for evaluating additional solvers in future gearbox lubrication studies.
\end{abstract}

\begin{highlights}
\item Benchmark study of four CFD solvers (OpenFOAM, Ansys Fluent, PreonLab, and MESHFREE) for predicting lubricant distribution in gearboxes.
\item Qualitative and quantitative validation of oil distribution simulations against experimental data from an FZG testbench.
\item Validation of grease simulation in a gearbox context using a Herschel-Bulkley SPH model.
\item Parametric study on the influence of gear speed and filling volume on lubrication deposition. 
\end{highlights}

\end{frontmatter}


\section{Introduction}

Gearboxes are fundamental components in a wide range of engineering applications, enabling mechanical power transfer between systems by either increasing torque while reducing speed or vice versa. Within Bosch's product portfolio, gearboxes are integral to systems such as eAxels, eBike drive units, and brake actuators. Many of these components not only involve complex rotating dynamics, but often operate in environments where multiphase flows are prevalent, further increasing the flow complexity. Capturing the interaction between rotating geometries and such multiphase flow phenomena with high fidelity remains a significant computational and experimental challenge.  

A typical gearbox consists of two or more gears enclosed within a housing, partially filled with lubricant to reduce friction, dissipate heat, and protect gear surfaces from wear. Maintaining the optimal lubricant volume is critical for both performance and durability: excess lubrication leads to increased churning losses, while insufficient lubrication accelerates gear wear, potentially leading to premature failure of components \cite{GearboxDesign1992}. Therefore, understanding and predicting the distribution of lubricants under realistic operating conditions is essential in the design and optimization of gearbox systems. 

Numerous experimental studies are conducted to study the lubricant distribution within gearboxes. W. Mauz et al. \cite{mauz-hvvsbub6m1988} performed a series of experiments ranging from single gear setups to two gear configurations, analysing the behaviour of lubricants under different operating conditions. Similarly, other researchers employed visualization techniques, such as high speed imaging and dye tracing methods, to capture oil splash patterns and quantify lubricant coverage on gear surfaces \cite{lubricants12080283, Timothy2005}. Despite these efforts, obtaining accurate and repeatable measurements remains difficult due to high rotational speeds and enclosed geometries.

To overcome the limitations of experimental methods, numerical simulations have become indispensable for analysing lubricant behaviour in gearboxes. However, accurately simulating these systems remains a complex task particularly due to the presence of rotating geometries. Such configurations introduce a considerable challenge in fluid flow simulations due to intricate dynamics and complex features that span a wide range of length scales, and few of these issues are addressed and discussed in \cite{Feng2021, Blais2022, Ruonan2024}. Different discretization methods and solvers are used to study the distribution of lubricant in gearboxes \cite{Groenenboom2019, Shadloo2016, refId0, LIU2017346, MASTRONE2020106496}. A comprehensive review of computational fluid dynamics (CFD) simulations addressing oil distribution in gearboxes, as well as other rotating systems such as pumps and bearings, is documented in the literature \cite{app10248810}. However, a systematic qualitative and quantitative comparison of different solvers applied to a predefined set of test cases is missing.

In addition to oil-lubricated systems, grease-lubricated gearboxes offer certain advantages, such as reduced leakage, and extended maintenance intervals. These benefits make grease a preferred lubricant in many enclosed or maintenance-constrained environments. The experimental work of Stemplinger in his Ph.D. thesis investigated the distribution of various greases under different operating conditions, highlighting the influence of viscosity and gear speed on coverage behaviour \cite{GreasePhD}. On the simulation side, Mastrone et al. employed OpenFOAM with a global remeshing approach to study grease distribution, validating their results both qualitatively and quantitatively \cite{GreaseSimulation1}. Hua et al. also performed qualitative comparisons between experimental and numerical results to evaluate the grease distribution within a gearbox \cite{GreaseSimulations2}. Despite these contributions, the number of studies focused on grease lubricated systems remains limited, particularly in terms of solver validation, parametric studies, and generalizability of findings across different setups.    

This study aims to fill this void by presenting a comparison of different solvers against a set of experiments for systems with complex rotating boundaries, specifically, the distribution of lubricant in a gearbox. To this end, the experimental data from  Hua Liu et al. \cite{lubricants6020047} is used to qualitatively validate the solvers for oil distribution within the gearbox. This is followed by a quantitative validation for the torque acting on the gear interacting with the oil \cite{mauz-hvvsbub6m1988}. Based on the validation results with oil as lubricants, the best performing solver is further qualitatively validated for grease distribution in a gearbox using the experimental results from Hua Liu et al. \cite{LiuGrease}. This solver is also used to investigate the influence of gear speed and lubricant filling volume on the amount of grease deposited on the gears.  

\section{Mathematical and Numerical Modelling}
\label{sec:mathematicalModeling}
We assume that the lubricant distribution in this study can be described as an incompressible, separated multiphase flow without froth or mist. Hence, the flow behaviour is described by the Navier-Stokes equations in the bulk domains \(\Omega\), and jump conditions at the interface \(\Sigma\) between the immiscible phases:
\begin{align}
    \nabla \cdot \textbf{u} &= 0 
    && \text{in } \Omega / \Sigma, \\
    \frac{\partial \rho \textbf{u}}{\partial t} + \nabla \cdot (\rho \textbf{uu}) 
    &= -\nabla p + \nabla \cdot \textbf{S} + \rho \mathbf{g} 
    && \text{in } \Omega / \Sigma.
\end{align}

Where \textbf{u} is the velocity, \(\rho\) is the density, \(p\) is the pressure, \(\mathbf{g}\) is the acceleration due to gravity, and \(\textbf{S} = \mu (\nabla \textbf{u} + (\nabla \textbf{u})^{T})\) is the viscous stress tensor, with \(\mu\) the dynamic viscosity.

 Fluid properties such as density and viscosity are assumed to be smooth functions of time and space within each phase but exhibit discontinuities at the interface, necessitating additional jump condition equations. The jump condition of any quantity \(\phi\) across the interface is defined as 
\begin{equation}
     \llbracket \phi \rrbracket = \lim_{\epsilon \to 0^+} [\phi (x_0 + \epsilon \textbf{n}_\Sigma) - \phi(x_0 - \epsilon \textbf{n}_\Sigma)]
 \end{equation}

 with \(\mathbf{n}_\Sigma\) denoting the surface normal pointing from liquid into the gas phase. Assuming no slip between the phases at the interface, the jump condition for the mass fluxes and for the stress at the interface, derived from the conservation of momentum, is given as 
 \begin{align}
    \llbracket \rho (\textbf{u} - \textbf{u}_{\Sigma}) \rrbracket \cdot \textbf{n}_{\Sigma} 
    &= \llbracket \dot{m} \rrbracket = 0
    && \text{on } \Sigma, \\
    \llbracket \rho \textbf{u}(\textbf{u} - \textbf{u}_{\Sigma}) + p\textbf{I} - \textbf{S} \rrbracket \cdot \textbf{n}_{\Sigma} 
    &= f_{\Sigma} 
    && \text{on } \Sigma.
 \end{align}
 
 Here, \(\dot{m}\) is the mass flux, \(\mathbf{u}_\Sigma\) is the interface velocity and \(f_{\Sigma}\) is the surface tension force.

In addition, for non-Newtonian fluids, e.g. some greases, the viscosity varies non-linearly with the applied stress. Therefore, the apparent viscosity \(\eta\) is used, defined as the ratio of shear stress \(\tau\) to strain rate \(\dot{\gamma}\) 
\begin{equation}
    \eta = \frac{\tau}{\dot{\gamma}} .
\end{equation}
Several models are available to describe the apparent viscosity, including Power-law, Bird-Carreau, Casson, Herschel-Bulkley, an overview can be found in \cite{RheologyGrease}. The general Herschel-Bulkley model, which is a combination of power-law model and Bingham plastic model, is given by 
\begin{equation}
    \eta = \frac{\tau_0}{\dot{\gamma}} + k\dot{\gamma}^{n-1}. 
\end{equation}
Here, \(\tau_0\) is the yield stress \([Pa]\), \(k\) is the consistency index \([Pa \cdot s^n]\), and \(n\) is the flow index \([dimensionless]\). Setting \(\tau_0 = 0\) reduces the equation to a power-law model, whereas \(n=1\) leads to the Bingham model. 

Multiple methods are available to discretize the above listed set of equations. Here, we apply the Volume of Fluid (VoF) method, Smoothed Particle Hydrodynamics (SPH), and the Finite Pointset Method (FPM). A brief overview is provided below.
\newpage
\subsection{Volume of Fluid}
\label{sec:VoF_Theory}
The Volume-of-Fluid method discretizes the one-field formulation of the two-phase flow equations on a mesh which decomposes the physical domain into finite volumes~\cite{CFDBasics,hirt1981volume}. To obtain the one-field formulation, a phase indicator is introduced to distinguish between the two phases and defines the volume fraction $\alpha$ as the integral over a control volume divided by its volume. In this study, the volume fraction is advected algebraically using the vanLeer scheme~\cite{Vof_vanLeer}. In order to sharpen the interface representation and ensure a geometrically consistent reconstruction of the interface with partially filled cells, the piecewise linear construction (PLIC) scheme is applied \cite{PLIC}. Surface tension effects are modelled using the continuous surface force (CSF) approach, originally proposed by Brackbill et al. \cite{BRACKBILL1992335}. For further details on the VoF method and the associated governing equations, the reader is referred to~\cite{Tomislav_VoF, VoF_Book}.

\begin{figure}[h!]
\centering
\begin{subfigure}{0.48\linewidth}
  \centering
  \small
  \resizebox{\linewidth}{!}{
\begingroup%
  \makeatletter%
  \providecommand\color[2][]{%
    \errmessage{(Inkscape) Color is used for the text in Inkscape, but the package 'color.sty' is not loaded}%
    \renewcommand\color[2][]{}%
  }%
  \providecommand\transparent[1]{%
    \errmessage{(Inkscape) Transparency is used (non-zero) for the text in Inkscape, but the package 'transparent.sty' is not loaded}%
    \renewcommand\transparent[1]{}%
  }%
  \providecommand\rotatebox[2]{#2}%
  \newcommand*\fsize{\dimexpr\f@size pt\relax}%
  \newcommand*\lineheight[1]{\fontsize{\fsize}{#1\fsize}\selectfont}%
  \ifx\svgwidth\undefined%
    \setlength{\unitlength}{231.5025654bp}%
    \ifx\svgscale\undefined%
      \relax%
    \else%
      \setlength{\unitlength}{\unitlength * \real{\svgscale}}%
    \fi%
  \else%
    \setlength{\unitlength}{\svgwidth}%
  \fi%
  \global\let\svgwidth\undefined%
  \global\let\svgscale\undefined%
  \makeatother%
  \begin{picture}(1,0.64130875)%
    \lineheight{1}%
    \setlength\tabcolsep{0pt}%
    \put(0,0){\includegraphics[width=\unitlength,page=1]{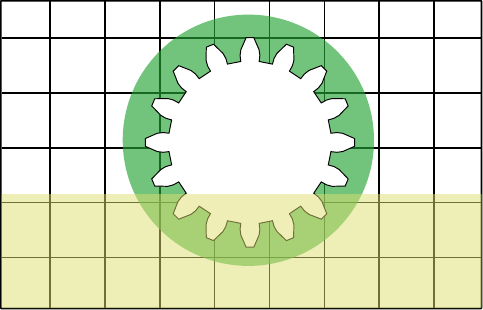}}%
    \put(0.35313732,0.33175104){\color[rgb]{0,0,0}\makebox(0,0)[lt]{\lineheight{1.25}\smash{\begin{tabular}[t]{l}Rotating Mesh\end{tabular}}}}%
    \put(0,0){\includegraphics[width=\unitlength,page=2]{Figures/AMI_Theory.pdf}}%
    \put(0.9413512,0.05302154){\color[rgb]{0,0,0}\rotatebox{89.715245}{\makebox(0,0)[lt]{\lineheight{1.25}\smash{\begin{tabular}[t]{l}Arbitrary Mesh Interface\end{tabular}}}}}%
    \put(0,0){\includegraphics[width=\unitlength,page=3]{Figures/AMI_Theory.pdf}}%
    \put(0.02908777,0.59476835){\color[rgb]{0,0,0}\makebox(0,0)[lt]{\lineheight{1.25}\smash{\begin{tabular}[t]{l}Stationary Mesh\end{tabular}}}}%
  \end{picture}%
\endgroup%
}
  \caption{Arbitrary Mesh Interface approach}
  \label{fig:VoF_AMI}
\end{subfigure}\hfill
\begin{subfigure}{0.48\linewidth}
  \centering
   \small
  \resizebox{\linewidth}{!}{
\begingroup%
  \makeatletter%
  \providecommand\color[2][]{%
    \errmessage{(Inkscape) Color is used for the text in Inkscape, but the package 'color.sty' is not loaded}%
    \renewcommand\color[2][]{}%
  }%
  \providecommand\transparent[1]{%
    \errmessage{(Inkscape) Transparency is used (non-zero) for the text in Inkscape, but the package 'transparent.sty' is not loaded}%
    \renewcommand\transparent[1]{}%
  }%
  \providecommand\rotatebox[2]{#2}%
  \newcommand*\fsize{\dimexpr\f@size pt\relax}%
  \newcommand*\lineheight[1]{\fontsize{\fsize}{#1\fsize}\selectfont}%
  \ifx\svgwidth\undefined%
    \setlength{\unitlength}{231.5025654bp}%
    \ifx\svgscale\undefined%
      \relax%
    \else%
      \setlength{\unitlength}{\unitlength * \real{\svgscale}}%
    \fi%
  \else%
    \setlength{\unitlength}{\svgwidth}%
  \fi%
  \global\let\svgwidth\undefined%
  \global\let\svgscale\undefined%
  \makeatother%
  \begin{picture}(1,0.64130875)%
    \lineheight{1}%
    \setlength\tabcolsep{0pt}%
    \put(0,0){\includegraphics[width=\unitlength,page=1]{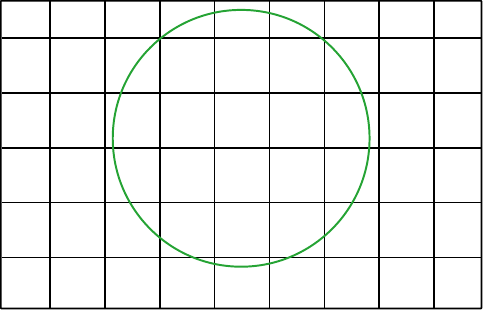}}%
    \put(0.34150225,0.32344027){\color[rgb]{0,0,0}\makebox(0,0)[lt]{\lineheight{1.25}\smash{\begin{tabular}[t]{l}Rotating Mesh\end{tabular}}}}%
    \put(0,0){\includegraphics[width=\unitlength,page=2]{Figures/Overset.pdf}}%
    \put(0.34002344,0.34492088){\color[rgb]{0,0,0}\makebox(0,0)[lt]{\lineheight{1.25}\smash{\begin{tabular}[t]{l}Rotating Mesh\end{tabular}}}}%
    \put(0,0){\includegraphics[width=\unitlength,page=3]{Figures/Overset.pdf}}%
    \put(0.02842924,0.59484073){\color[rgb]{0,0,0}\makebox(0,0)[lt]{\lineheight{1.25}\smash{\begin{tabular}[t]{l}Stationary Mesh\end{tabular}}}}%
    \put(0,0){\includegraphics[width=\unitlength,page=4]{Figures/Overset.pdf}}%
    \put(0.88553572,0.092561){\color[rgb]{0,0,0}\rotatebox{89.974369}{\makebox(0,0)[lt]{\lineheight{1.25}\smash{\begin{tabular}[t]{l}Overlapping rotating\\and stationary mesh\end{tabular}}}}}%
  \end{picture}%
\endgroup%
}
  \caption{Overset approach}
  \label{fig:OF_VoF_overset}
\end{subfigure}

\caption{Schematic illustration of domain decomposition and mesh rotation approaches tested with Volume of Fluid method. Note: The white region in the domain represents air phase, while the yellow region represents the oil phase.}
\label{fig:Vof_MeshTheory}
\end{figure}

In addition to the multiphase nature of lubrication, the system under consideration involves rotating geometries.
To accurately capture such complex moving boundaries with mesh based methods, multiple approaches are available. In this study, two types are investigated.
The Arbitrary Mesh Interface (AMI) method enables the domain to be partitioned into stationary and rotating regions, facilitating smooth interpolation of field variables across these non-conformal mesh interfaces, see \cref{fig:VoF_AMI}. This method is similar to a sliding mesh approach.
In contrast, the overset approach (also known as chimera grids) overlays the rotating region onto the stationary region, shown as the green mesh in \cref{fig:OF_VoF_overset}, with flow field data interpolated within the overlapping regions to ensure continuity. For more detailed information on these mesh rotation methods, the reader is referred to \cite{FARRELL20092632, FARRELL201189, AMIJasak, Benek1983}. This study employs OpenFOAM v2212 with the interFoam solver \cite{OpenFOAMcode} and Ansys-Fluent 2302 \cite{ansysfluentuser2023r2,ansysfluenttheory2023r2}.

\subsection{Smoothed Particle Hydrodynamics}
\label{sec:SPH_Theory}
\begin{figure}[b!]
  \centering
  \def\svgwidth{0.99\textwidth}
  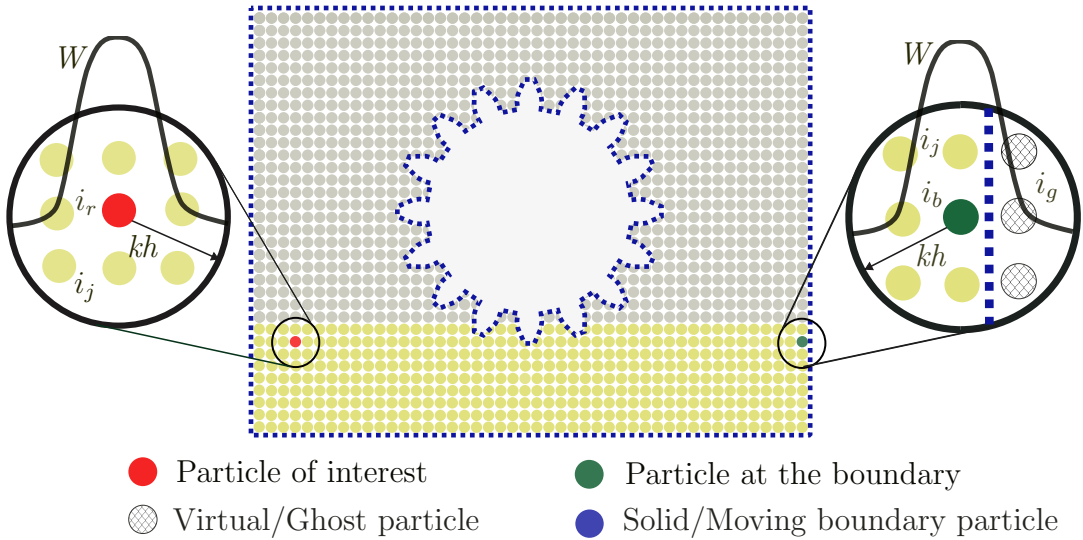
  \captionof{figure}{Schematic representation of domain decomposition and particle distribution in Smoothed Particle Hydrodynamics: Left inset shows reference particle '\(i_r\)' (red) has its properties influenced by the neighbour particles '\(i_j\)' within the domain of influence 'kh'. Right inset illustrates the interaction for a boundary particle '\(i_b\)' (green).}
  \label{fig:SPH_Theory}
\end{figure}

Smoothed Particle Hydrodynamics (SPH), first introduced in \cite{SPH_Initial1, SPH_Initial2}, is a Lagrangian method to discretize the fluid in the computational domain into a finite number of particles, each carrying physical properties such as mass, velocity, and pressure \cite{SPH_Review}. Physical quantities are computed using weighted interpolations of neighbouring particle properties. The interaction between particles is governed by kernel functions, with only particles within the influence domain contributing to its behaviour. A schematic representation of this method is illustrated in \cref{fig:SPH_Theory}.

The kernel function, \(W\), must satisfy the normalization condition, along with additional properties such as positivity, compact support, and smoothness. Commonly used kernel functions in SPH include quartic spline, cubic spline, Gaussian, quadratic, double cosine \cite{SPH_DoubleCosineKernel}. In this paper, the cubic spline kernel function is employed. For a comprehensive discussion on kernel properties and various kernel functions, reader is referred to \cite{SPH_KernelFunctionOverview}.
Solid boundaries in SPH are often modelled using ghost particle method \cite{Sph_GhostParticle1, Sph_GhostParticle2, Sph_GhostParticle3}. Physical properties such as velocity, temperature and pressure are attributed to these particles from which the approximate properties of the fluid particles are obtained. To avoid the penetration of fluid particles into the solid boundary, an artificial repulsive force calculated by an analogy with the Lennard-Jones molecular force is used \cite{SPH_LennardForce}. A detailed information on the other methods used to model the boundary in SPH can be found in \cite{SPHBasics}.    

Various methods exist for modelling surface tension, including cohesive pressure model \cite{SPH_ST1}, inter-particle force model \cite{SPH_ST2, SPH_ST3}, and continuum surface force \cite{SPH_ST4}. In this study, for a single phase  (oil or grease) discretization, we employ the potential force model---an extension of pairwise force model \cite{SPH_Pairwise}---to capture surface tension effects. In this model, additional particle–particle interaction forces are incorporated into the momentum equation to account for both surface tension and fluid–solid interactions. 


Although both the lubricant and air phases can be discretized, a key advantage of this Lagrangian approach involves discretizing only the heavier phase, significantly reducing the simulation time. Furthermore, the mesh-free nature of SPH makes it particularly suitable for complex rotating geometries, such as gears, eliminating challenges related to mesh distortion and interpolation errors. In this study, PreonLab version 5.3.2 is employed \cite{preonlab}.

\subsection{Finite Pointset Method}
Finite Pointset Method (FPM), similar to SPH method, is a Lagrangian approach where the computational domain is discretized into a finite set of unstructured points, each carrying physical properties such as velocity, pressure, and density. However, unlike SPH, differential operators are approximated using a least squares regression over neighbouring points \cite{FPM_Theory001}. 

To solve the Navier-Stokes equations, the momentum and mass conservation equations are solved in two separate steps. The particle positions and intermediate velocities are first updated explicitly, followed by a pressure correction step to enforce the incompressibility constraint. The pressure field is obtained by solving a pressure Poisson equation, using Chorin-based projection method \cite{ChorinProjectionMethod}, and the velocity field is subsequently corrected using the computed pressure gradient \cite{FPM_Theory003}. 


As mentioned above, the pressure gradient and viscous terms are evaluated using a weighted least squares approach, where neighbouring particles within a defined interaction radius are used to locally approximate spatial derivatives. A Gaussian weight function is employed to ensure that closer neighbours have stronger influence on the approximation, improving accuracy and numerical stability. 

For surface tension, the Boundary Force Model, also referred to as the Direct Surface Tension Model is employed. The surface tension is incorporated directly as a boundary condition on relevant particles - specifically free surface particles in single phase simulations and is treated as a direct surface force, eliminating the need to convert it into an equivalent volume force over a finite thickness \cite{FPM_Theory002}. 

In this work, the FPM method is used to discretize only the heavy phase (oil) in the domain. Although it is possible to discretize all phases in a multiphase system, the key advantage lies in resolving only the relevant fluid phase, significantly reducing computational effort. Similar to SPH, the absence of mesh makes FPM well-suited for complex rotating geometries, such as gears and bearings. MESHFREE version 2023.05.0 is used to test this approach \cite{Meshfree}.

\subsection{Benchmarked tools}
\label{sec:solver-overview}
To investigate the suitability of the three numerical methods described above, four different solvers were used, as summarized in \cref{tab:solver-overview}. The solvers are chosen based on their common usage in either industry or academia or both and their future potential. For all solvers, we use the recommended settings.

\begin{table}[bp!]
    \begin{tabular}{ p{1.5cm}p{1.8cm}p{2cm}p{1.8cm}>{\raggedright\arraybackslash}p{2.2cm} p{2.5cm}}
     \toprule
     Solver &  Publisher & License & Method & Discretization & Mesh-motion  \\
    \midrule
     interFoam & OpenCFD Ltd & Gnu GPL v3 & algebraic VoF &unstructured hexahedral &  AMI/sliding\\
     Fluent &Ansys Inc. & commercial & algebraic VoF & unstructured polyhedral &  AMI/sliding overset \\  
     PreonLab & Fifty2 & commerical  & SPH & only liquid& \\
     Meshfree & Fraunhofer SCAI* & commerical  & FPM & only liquid &\\
    \bottomrule
    \end{tabular}
    \caption{Overview of benchmarked solvers.}
    \label{tab:solver-overview}
    \footnotesize{$^*$ Fraunhofer Institute for Algorithms and Scientific Computing}
\end{table}

All present solvers offer the additional capability of adaptive resolution refinement; however, this feature is not utilized in the present work to maintain consistency across simulations. Furthermore, PreonLab offers a GPU solver, which shortens simulation time by approximately a factor of six but was not used for the comparison. Ansys-Fluent has shared first GPU capabilities, but mesh motion is not included yet. Consequently, all simulations in the validation study are conducted on an \textit{Intel (R) Xeon (R) Gold 6342} based Bosch High Performance Computing (HPC) Cluster. Geometry generation is carried out using FreeCAD ~\cite{freecad2024} for OpenFOAM, PreonLab, and MESHFREE, while Ansys DesignModeler is used for generating geometries compatible with Ansys-Fluent Meshing. For OpenFOAM related simulations, version v2212 is used in combination with cfMesh \cite{cfMesh} for mesh generation. Ansys-Fluent v2302 is employed along with Ansys-Fluent Mesher v2302 for grid generation~\citep{ansysmeshinguser2023r2}. Post-processing of OpenFOAM, Ansys-Fluent, and MESHFREE results is performed using ParaView \cite{Paraview}, whereas the built-in post-processing tools of PreonLab are used for its respective results.

\section{Oil distribution in gearbox} 
\label{sec:results1gear}
\begin{figure}[tb!]
	\centering
	\includegraphics[width=0.38\linewidth]{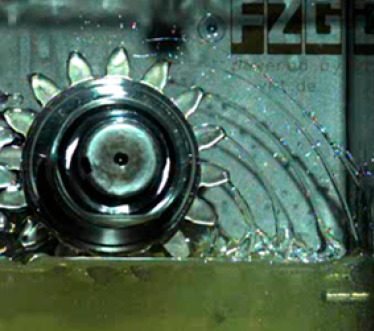}
	\caption{Experimental oil distribution used as a reference for each solver. This picture is taken from the work of Hua Liu et al. \cite{lubricants6020047}.  }
	\label{fig:FZG_Experimental_Expected}
\end{figure}

In this study, two distinct setups are used for validation: one for qualitative analysis and the other for quantitative evaluation. The qualitative setup focuses on observing and comparing flow patterns around the gear and casing, see \cref{fig:FZG_Experimental_Expected}. The visualization focuses exclusively on the right side of the gearbox, and the effects of the driving gear are ignored. In contrast, the quantitative setup is used to compare the torque acting on a single gear due to the lubricant sump. 

\subsection{Qualitative Comparison}
\label{sec:singlegearCase1}

\begin{figure}[htbp!]
  \centering
  \begin{subfigure}[t]{0.40\textwidth}
    \centering
    \includegraphics[width=\linewidth]{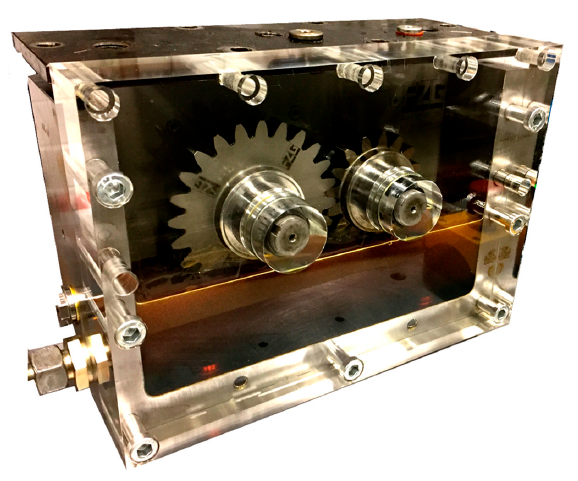}
    \label{fig:FZG_OriginalSetup}
  \end{subfigure}
  \hfill
  \begin{subfigure}[t]{0.54\textwidth}
    \centering
    \def\svgwidth{\linewidth}
    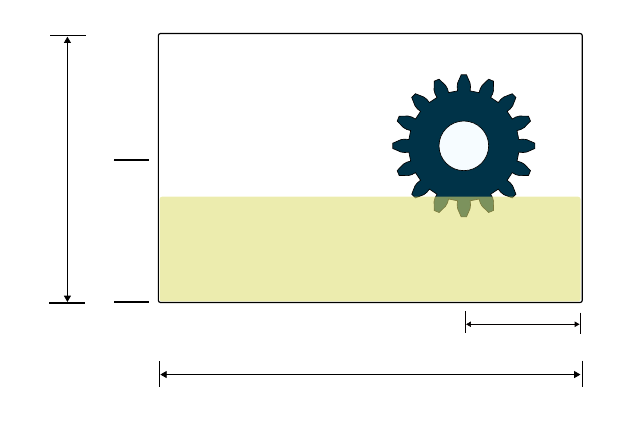
    \label{fig:FZG_UsedSetup}
  \end{subfigure}
  \caption{Illustration of FZG setup. Left: FZG setup used in the experiments. This picture is taken from the work of Hua Liu et al. \cite{lubricants6020047}. Right: Simulation setup with only pinion gear.}
  \label{fig:FZGSetup}
\end{figure}

This section uses the setup and experimental results from Hua Liu et al. \cite{lubricants6020047}. In their experiments, the authors utilized a housing with dimensions $L =$ 210 mm (length), $H =$ 175 mm (height), $W =$ 56 mm (width), containing two rotating gears. To reduce simulation complexity, only the pinion is considered, rotating at 240 RPM, see \cref{fig:FZGSetup}. The geometric properties of the pinion are given in \cref{tab:GeometryParameter}. The pinion is placed at a distance $X_c = 59.25 $ mm from the right wall and $Y_c = 87.5$ mm from the bottom wall of the housing. The level of lubricant filling, denoted by $e$, corresponds to a vertical height of 9 mm measured from the tip of the gear to the surface of the lubricant. 

\begin{figure}[bt!]
    \centering
    \begin{subfigure}{0.3\textwidth}
    \centering
    \includegraphics[width=\textwidth]{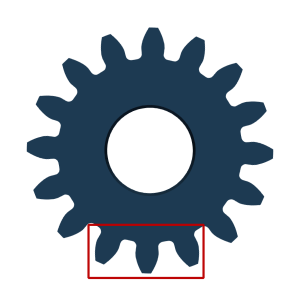}
    \label{fig:FZGSetup_MeshLocation}
    \end{subfigure}
    \\
    \begin{subfigure}{0.38\textwidth}
        \centering
        \includegraphics[width=\textwidth]{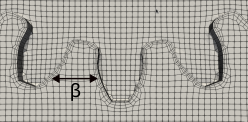}
        \label{fig:OF_FZG_MediumResolution}
        \caption{OpenFOAM with \(\beta = \text{7 cells}\)}
    \end{subfigure} 
    \begin{subfigure}{0.40\textwidth}
        \centering
        \includegraphics[width=\textwidth]{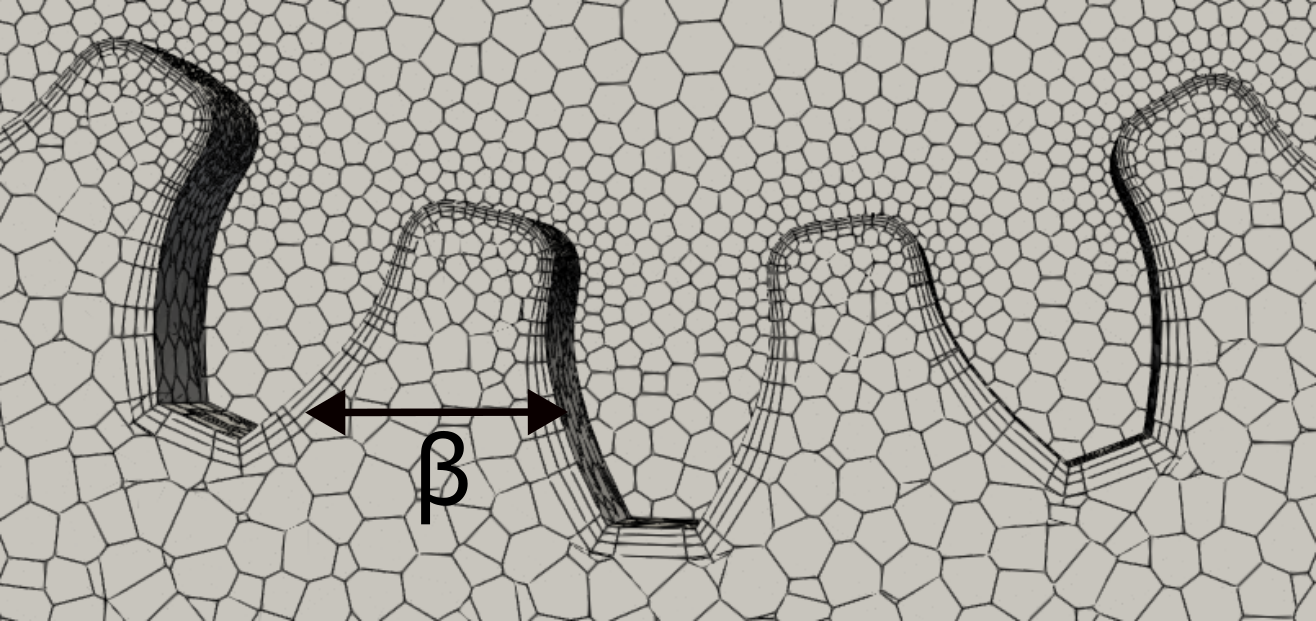}
		\label{fig:AF_FZG_MediumResolution}
        \caption{Ansys-Fluent with \(\beta = \text{7 cells}\)}
    \end{subfigure}
    \\
    \begin{subfigure}{0.38\textwidth}
        \centering
        \includegraphics[width=\textwidth]{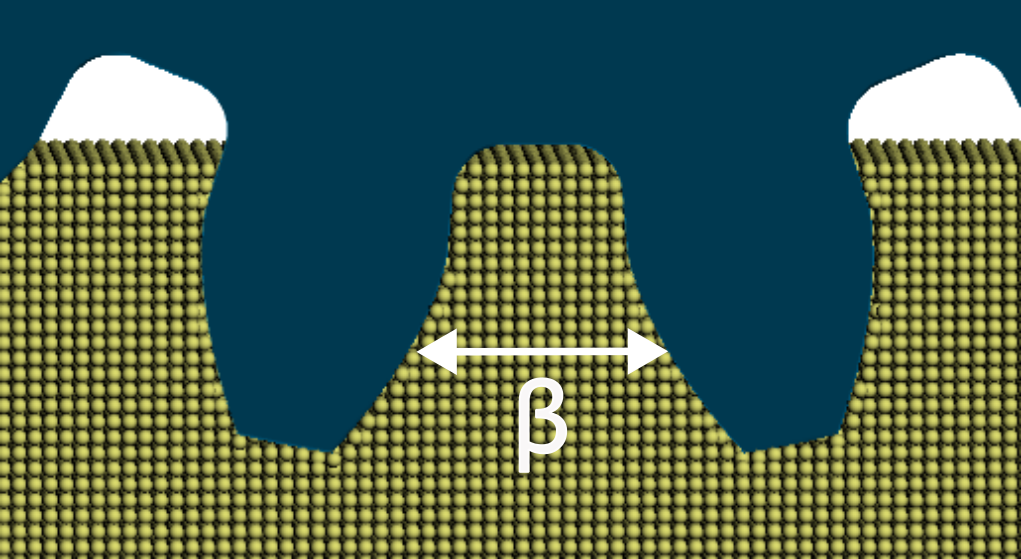}
        \label{fig:PL_Fine_ParticleResolution}
        \caption{PreonLab with \(\beta = \text{7 particles}\)}
    \end{subfigure}
    \begin{subfigure}{0.40\textwidth}
        \centering
        \includegraphics[width=\textwidth]{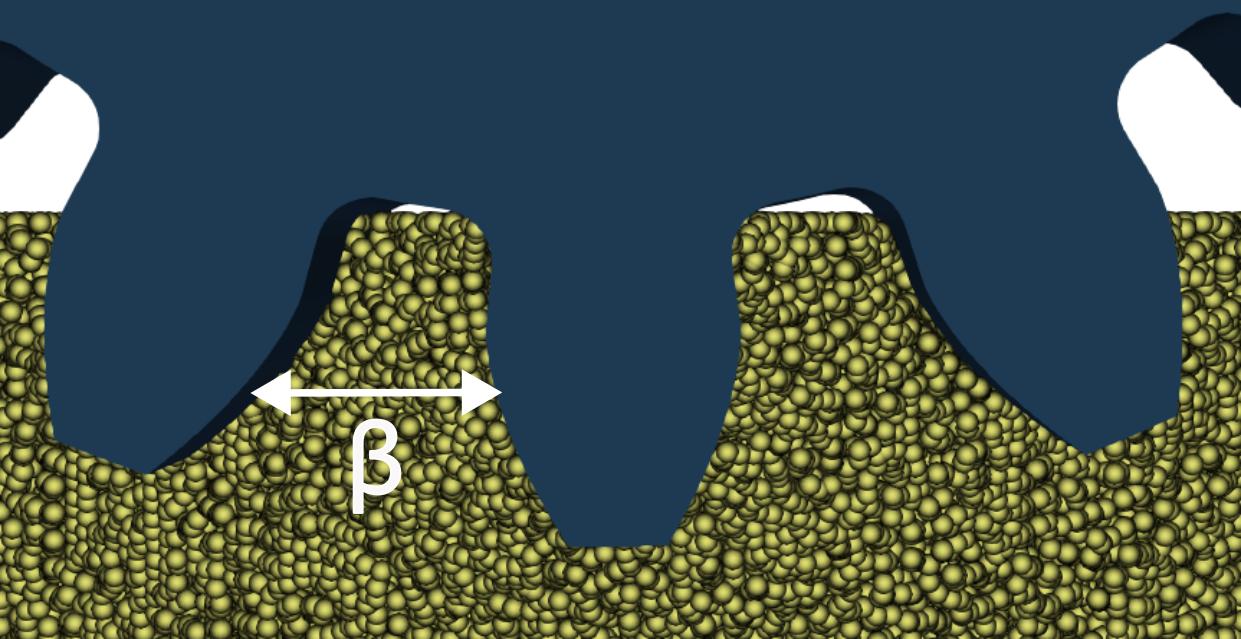}
        \label{fig:PL_FineParticles_ParticleResolution}
        \caption{MESHFREE with \(\beta = \text{12 particles}\)}
    \end{subfigure}
    \caption{Visualization of mesh and particle resolution for intermediate discretization level across different solvers. Top: Pinion gear employed in this study, with the red box highlighting the area where the mesh or particles are illustrated in the subsequent images. Bottom: Mesh or particle resolution in each solver.}
    \label{fig:FZG_MeshOrParticleResolution}
\end{figure}

The properties of the lubricant are summarized in \cref{tab:LubricantParameter}. To allow an objective tool comparison, we define the resolution based on the discretization particles or cells between two adjacent teeth, see \cref{fig:FZG_MeshOrParticleResolution}.

\begin{table}[h!]
    \RawFloats
    \centering
    \begin{minipage}[t]{0.45\textwidth}
        \vspace{0pt}  
        \centering
        \begin{tabular}{@{}ll@{}}
            \toprule
            Parameter                   & Value \\ \midrule
            Number of teeth                  & 16            \\
            Module in $mm$       & 4.5            \\
            Pressure angle in $degrees$              & 20            \\
            Pitch diameter in $mm$              & 118.4         \\
            \bottomrule
        \end{tabular}
        \caption{Geometric properties of the pinion.}
        \label{tab:GeometryParameter}
    \end{minipage}
    \hfill
    \begin{minipage}[t]{0.45\textwidth}
        \vspace{0pt}
        \centering
        \begin{tabular}{@{}ll@{}}
            \toprule
            Parameter                   & Value \\ \midrule
            Density in $kg/m^3$                   & 855           \\
            Kinematic viscosity in $mm^2/s$       & 32            \\
            Contact angle in $degrees$              & 45            \\
            Surface tension in $N/m$              & 0.032         \\
            \bottomrule
        \end{tabular}
        \caption{Properties of the lubricant.}
        \label{tab:LubricantParameter}
    \end{minipage}
\end{table}


\subsubsection{Results}
\label{sec:singlegGearQualitative}
In the following, first a resolution study is discussed for each solver, followed by a direct comparison using the best-performing setup identified for each solver.

\textbf{OpenFOAM}

A resolution analysis using hexahedral meshes is performed with three configurations: 4, 7, and 10 cells between two consecutive gear teeth. The simulations are carried out using the interFoam solver, which implements an algebraic Volume of Fluid (VoF) method; see \cref{sec:VoF_Theory} for further details.

As described in \cref{sec:solver-overview}, post-processing is conducted using ParaView. To visualize the oil distribution, both threshold and contour are applied to the alpha field. A threshold filter with a minimum value of 0.5 and a maximum value of 1.0 is used to visualize the oil reservoir. A contour is defined at an alpha value of 0.1 to highlight thin fluid filaments (fingers) along the gear teeth, ensuring that only regions with significant fluid presence are displayed.

\begin{figure}[tbp!]
	\centering
	\begin{subfigure}{0.32\textwidth}
		\centering
		\includegraphics[width=\textwidth]{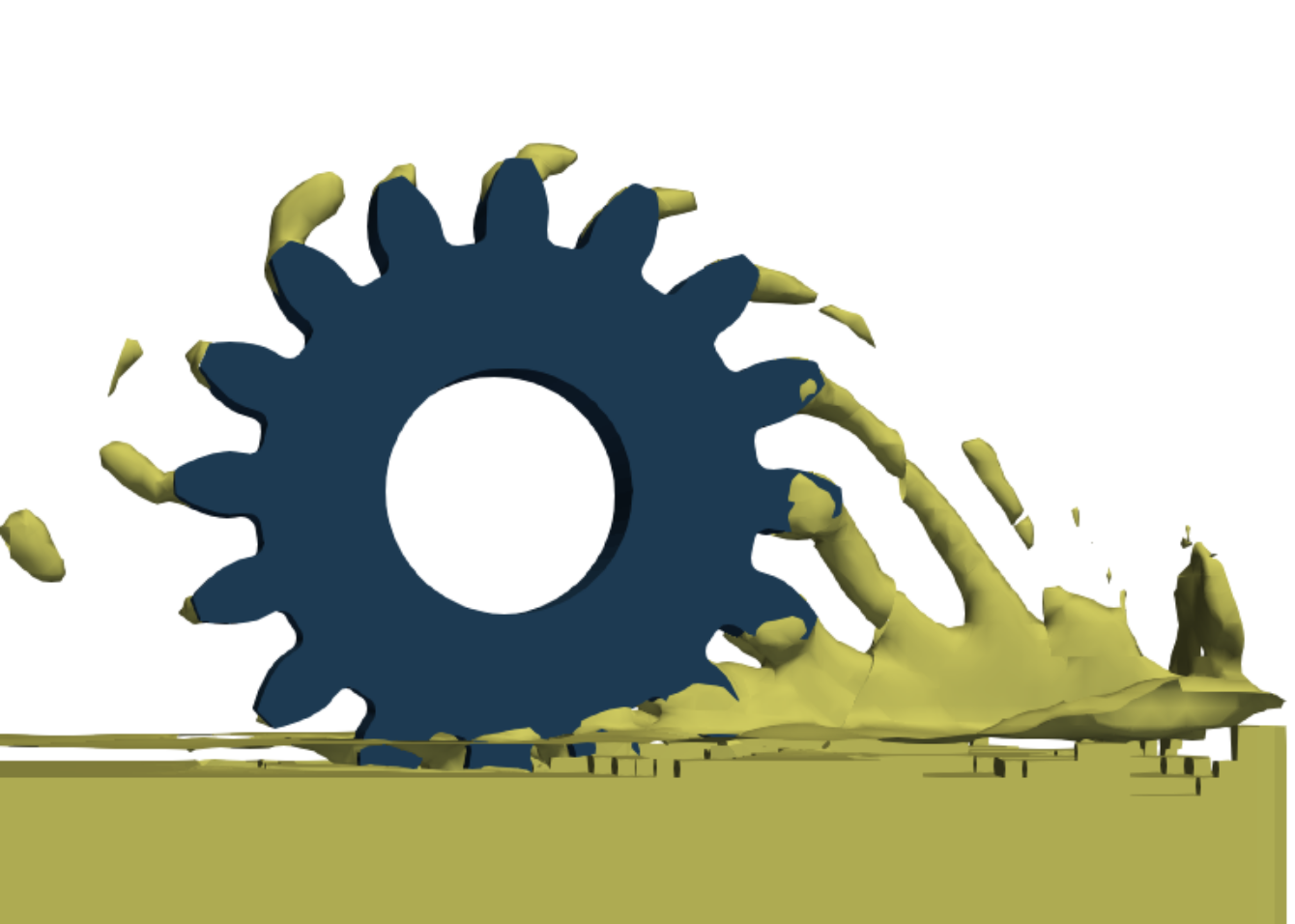}
		\caption{Coarse resolution\newline\hspace*{1.5em}Time taken = 1.49 hours}
		\label{fig:OF_FZG_CoarseResolution_Result}
	\end{subfigure}
	\begin{subfigure}{0.32\textwidth}
		\centering
		\includegraphics[width=\textwidth]{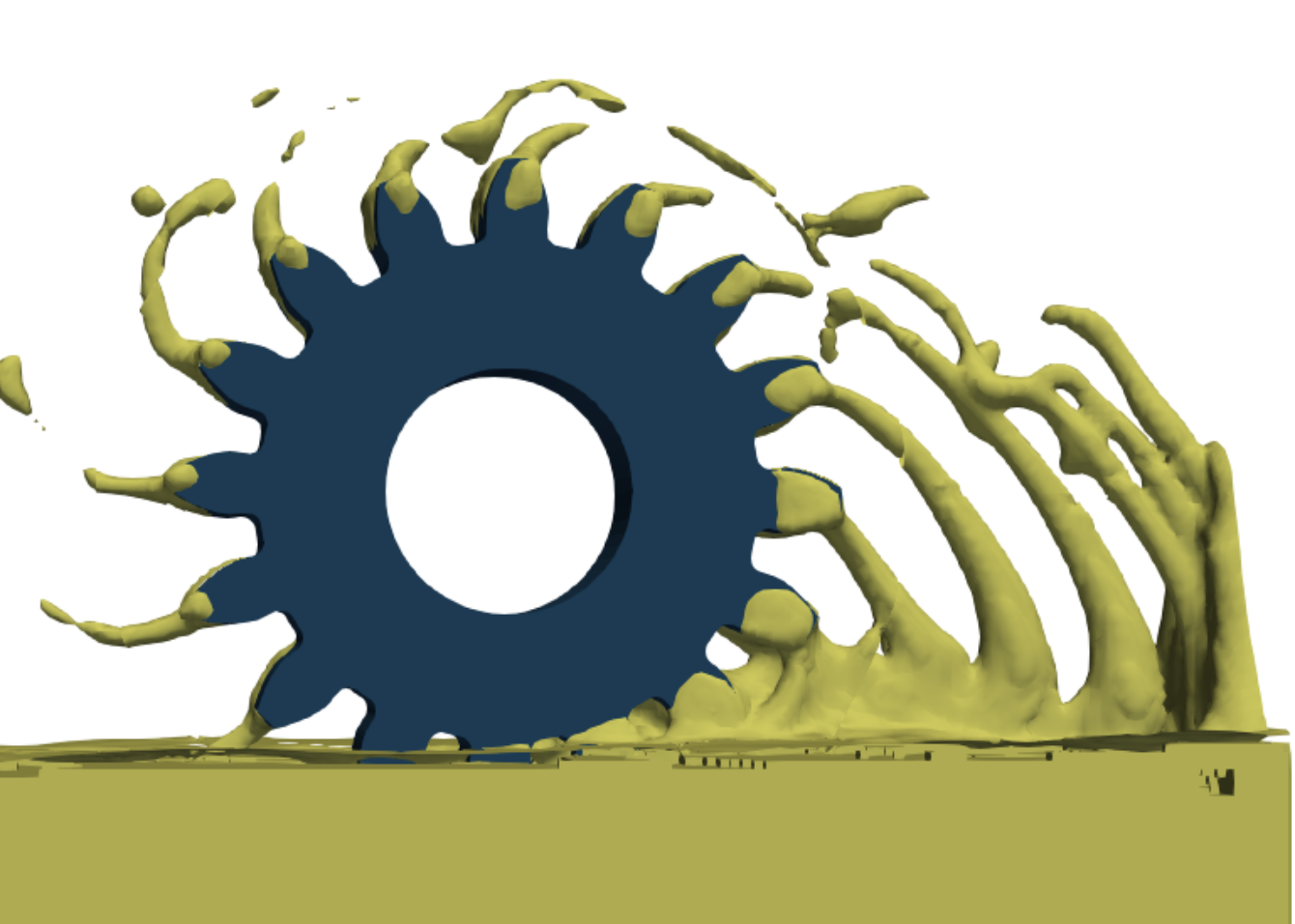}
		\caption{Medium resolution\newline\hspace*{1.5em}Time taken = 24.43 hours}
		\label{fig:OF_FZG_MediumResolution_Result}
	\end{subfigure}
	\begin{subfigure}{0.32\textwidth}
		\centering
		\includegraphics[width=\textwidth]{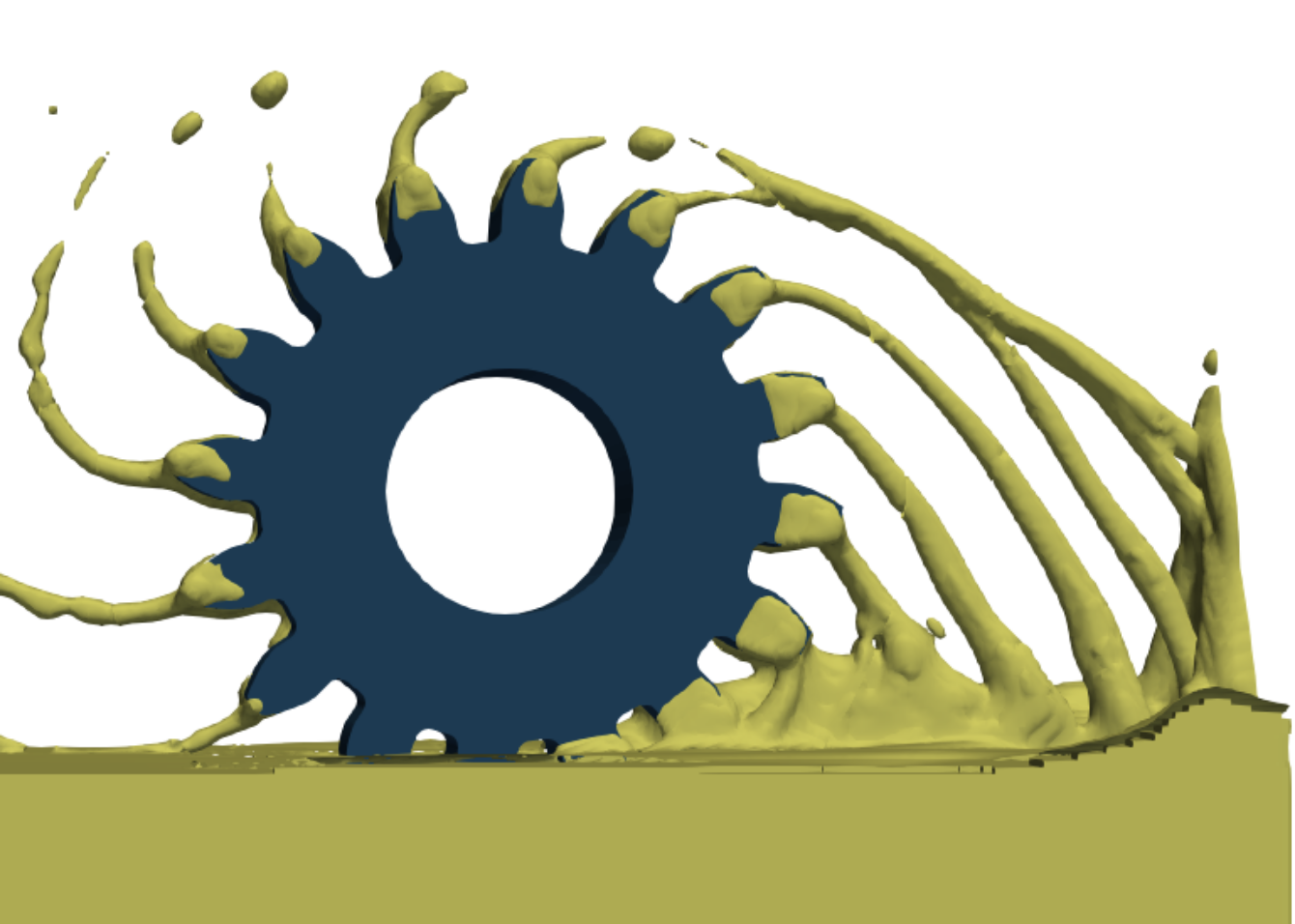}
		\caption{Fine resolution\newline\hspace*{1.5em}Time taken = 26.66 hours}
		\label{fig:OF_FZG_FineResolution_Result}
	\end{subfigure}
	\caption{Comparison of fluid distribution at different mesh resolutions using \textbf{OpenFOAM} with AMI/sliding mesh approach. }
	\label{fig:FZG_OpenFOAM_MeshStudy}
\end{figure}

\begin{figure}[tb!]
	\centering
	\includegraphics[width=0.7\textwidth]{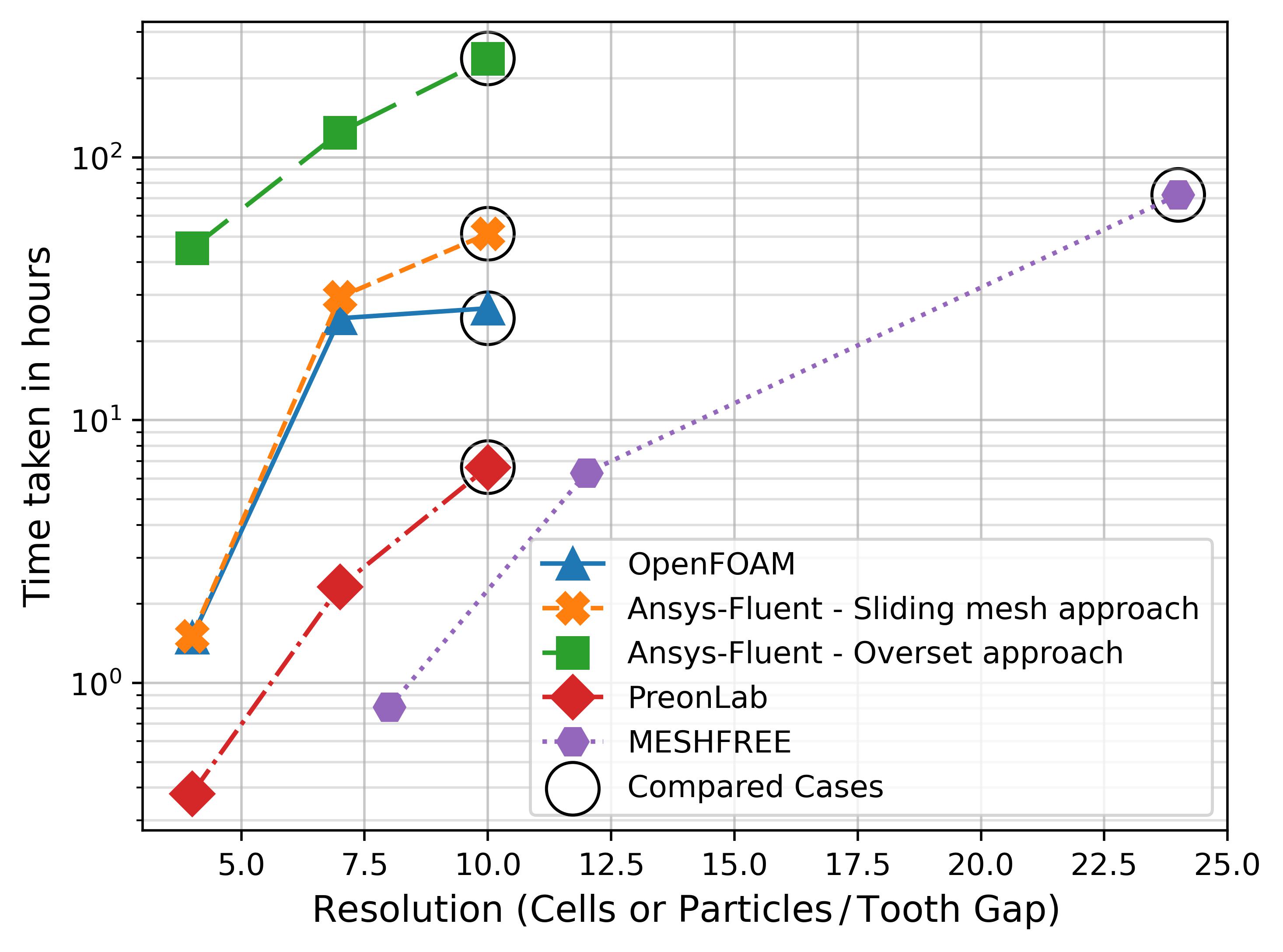}
	\caption{Plot showing the resolution and time taken for each resolution in each solver.}
	\label{fig:FZG_AllSoftware}
\end{figure}

\Cref{fig:FZG_OpenFOAM_MeshStudy} illustrates the oil distribution obtained with different mesh resolutions. At the coarsest resolution (4 cells across the gear tooth width), oil fingers detach as early as the fourth tooth above the oil sump, indicating under-resolution. With a refined mesh of 7 cells, detachment shifts to the fifth tooth, reflecting improved accuracy. Further refinement to 10 cells yields a similar detachment location; however, the oil distribution appears more distinctly resolved compared to the 7-cell case. Given the improved accuracy and only marginally higher computational cost, the 10-cell configuration was selected for all subsequent simulations. The computational cost comparison is provided in \cref{fig:FZG_AllSoftware} and discussed in detail later. It should be noted that the overset approach in OpenFOAM was excluded from this study, as it failed to achieve mass conservation.  

\textbf{Ansys-Fluent} 

In Ansys-Fluent, both the sliding mesh and overset mesh approaches are evaluated. Computational grids are generated using the Fluent Mesher, with polyhedral cells selected as the default meshing strategy. Post-processing is performed using ParaView with visualization settings consistent with those used in OpenFOAM. Three mesh resolutions are analysed, corresponding to 4, 7, and 10 cells between two consecutive gear teeth.

\begin{figure}[bt!]
	\centering
	\begin{subfigure}{0.30\textwidth}
		\centering
		\includegraphics[width=\textwidth]{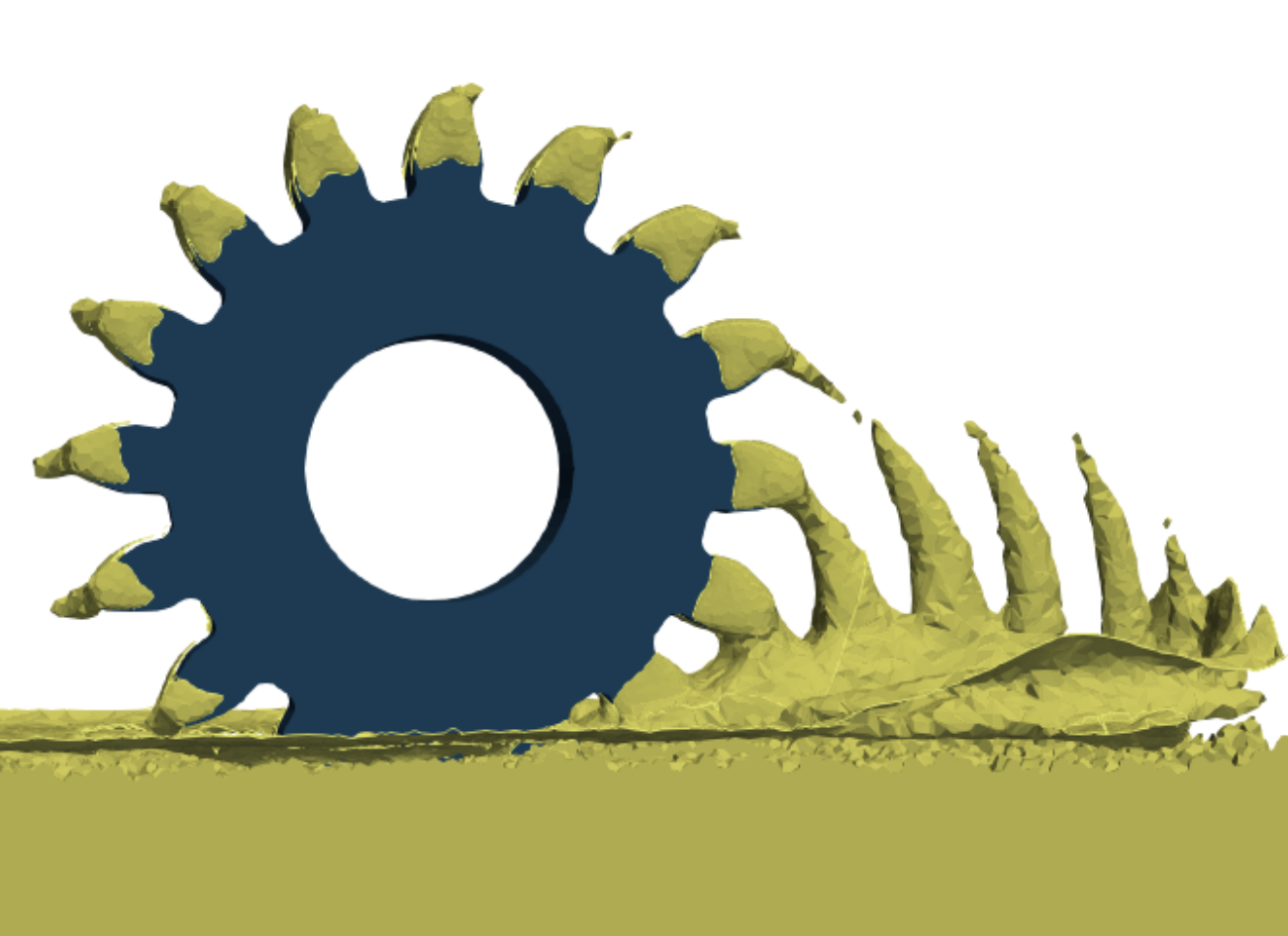}
		\caption{Coarse resolution\newline\hspace*{1.5em}Time taken = 1.51 hours}
		\label{fig:AF_FZG_CoarseResolution_Result}
	\end{subfigure}
	\begin{subfigure}{0.30\textwidth}
		\centering
		\includegraphics[width=\textwidth]{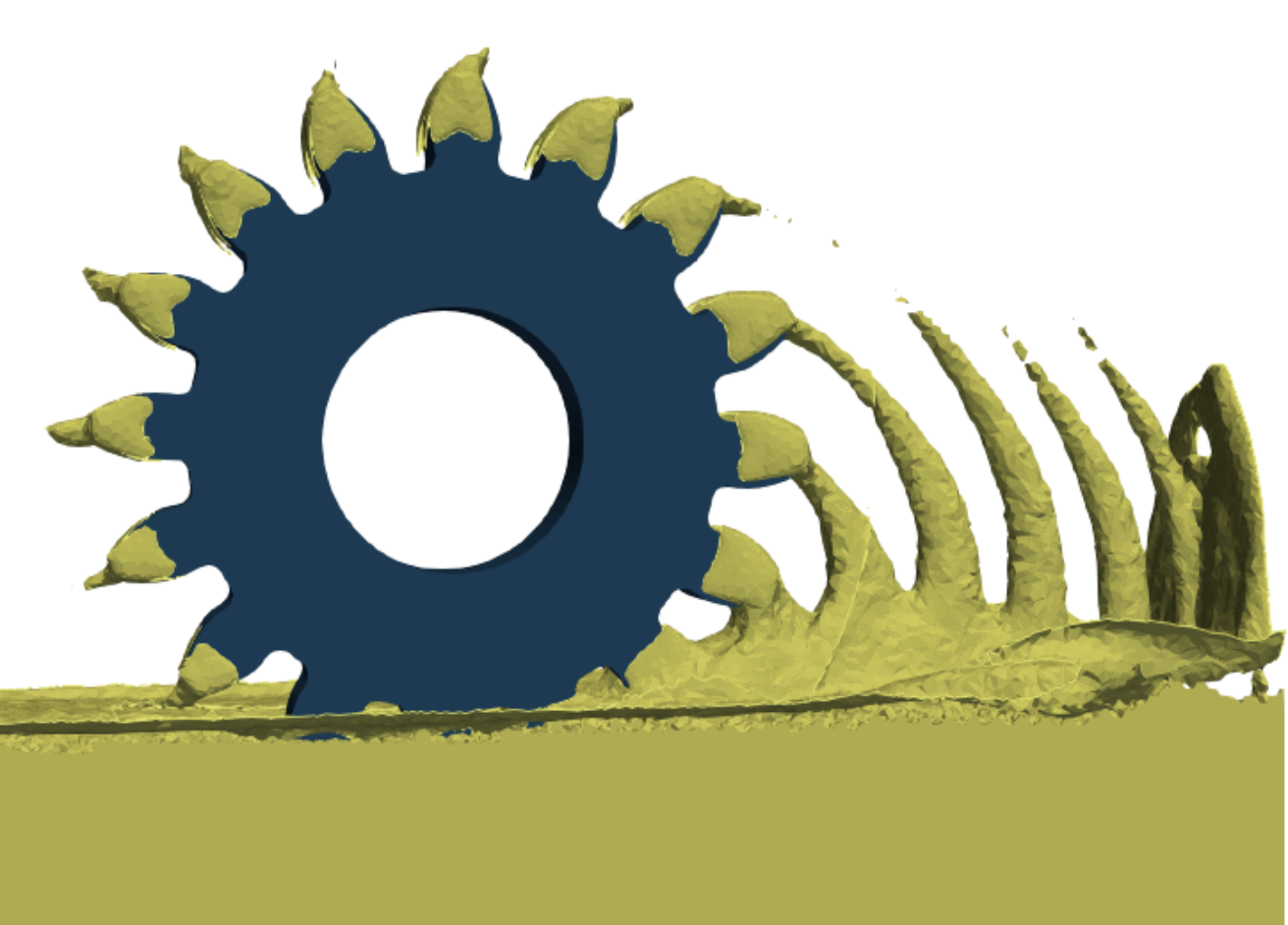}
		\caption{Medium resolution\newline\hspace*{1.5em}Time taken = 29.43 hours}
		\label{fig:AF_FZG_MediumResolution_Result}
	\end{subfigure}
	\begin{subfigure}{0.30\textwidth}
		\centering
		\includegraphics[width=\textwidth]{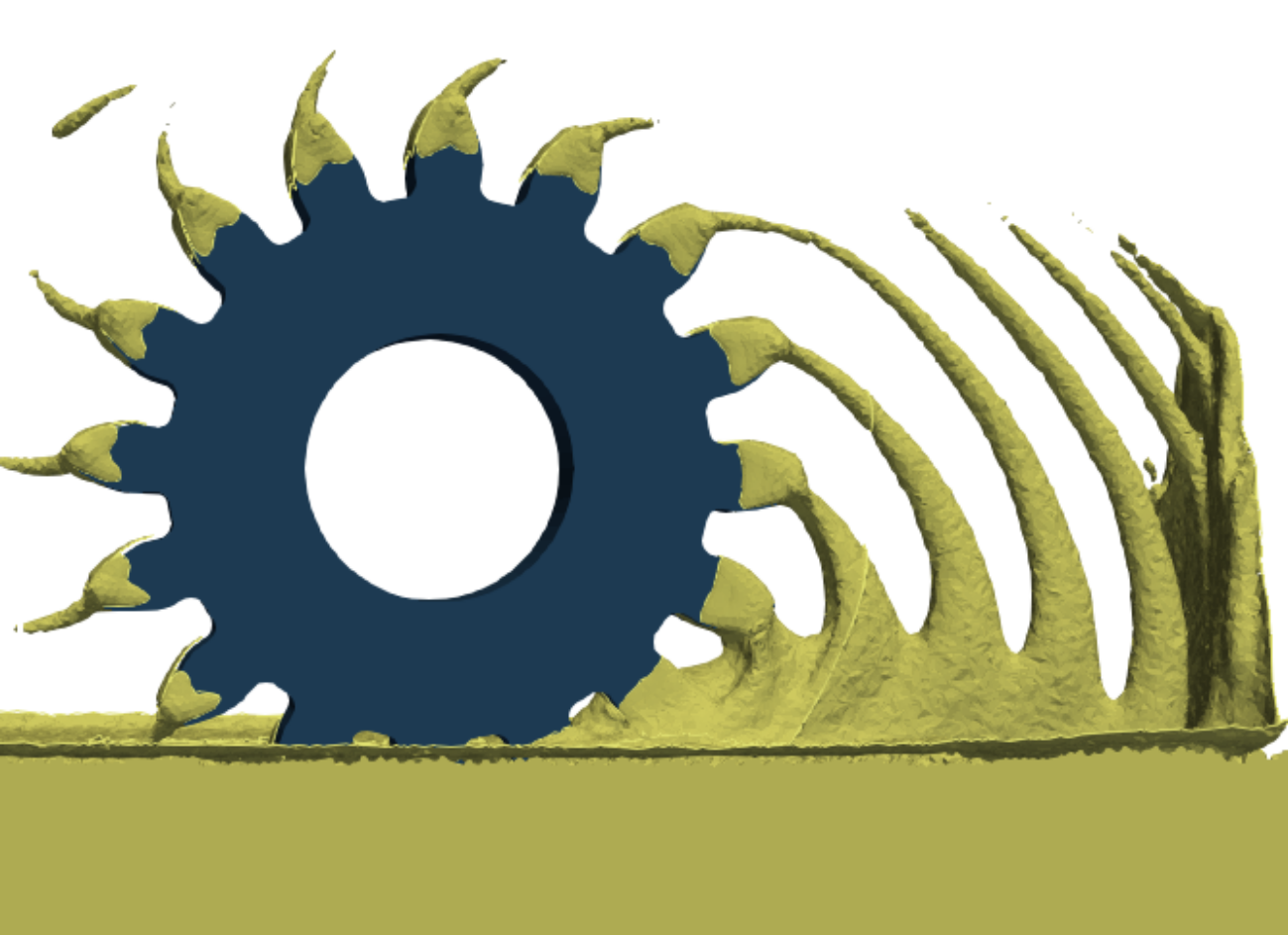}
		\caption{Fine resolution\newline\hspace*{1.5em}Time taken = 51.22 hours}
		\label{fig:AF_FZG_FineResolution_Result}
	\end{subfigure}
	\\
	\caption{Comparison of fluid distribution at different mesh resolutions with \textbf{Ansys-Fluent} using AMI/sliding mesh approach. }
	\label{fig:AF_FZG_Results}
\end{figure}

As shown in \cref{fig:AF_FZG_Results}, for the sliding mesh approach, resolution clearly affects the detachment location. At the coarsest resolution (4 cells), detachment occurs at the third tooth above the oil sump. With 7 cells, detachment shifts to the fourth tooth, and at the finest resolution (10 cells), it occurs at the fifth tooth. Further mesh refinement does not lead to significant changes in the overall distribution pattern, while also increasing computational cost substantially. Therefore, the 10-cell configuration is selected for subsequent comparisons.

\begin{figure}[htbp!]
	\centering
	\begin{subfigure}{0.32\textwidth}
		\centering
		\includegraphics[width=\textwidth]{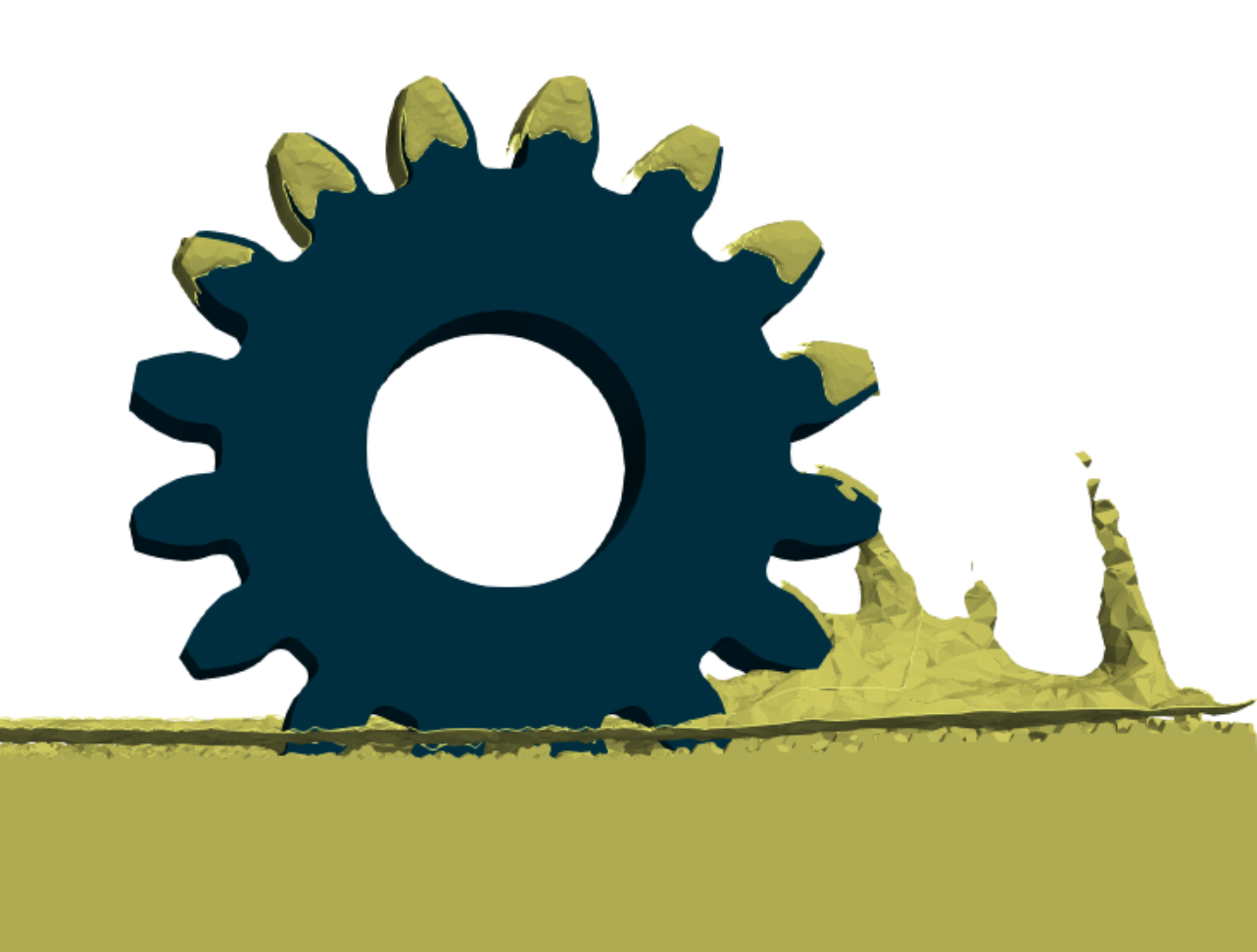}
		\caption{Coarse resolution\newline\hspace*{1.5em}Time taken = 45.13 hours}
		\label{fig:OF_FZG_OversetCoarseResolution_Result}
	\end{subfigure}
	\begin{subfigure}{0.32\textwidth}
		\centering
		\includegraphics[width=\textwidth]{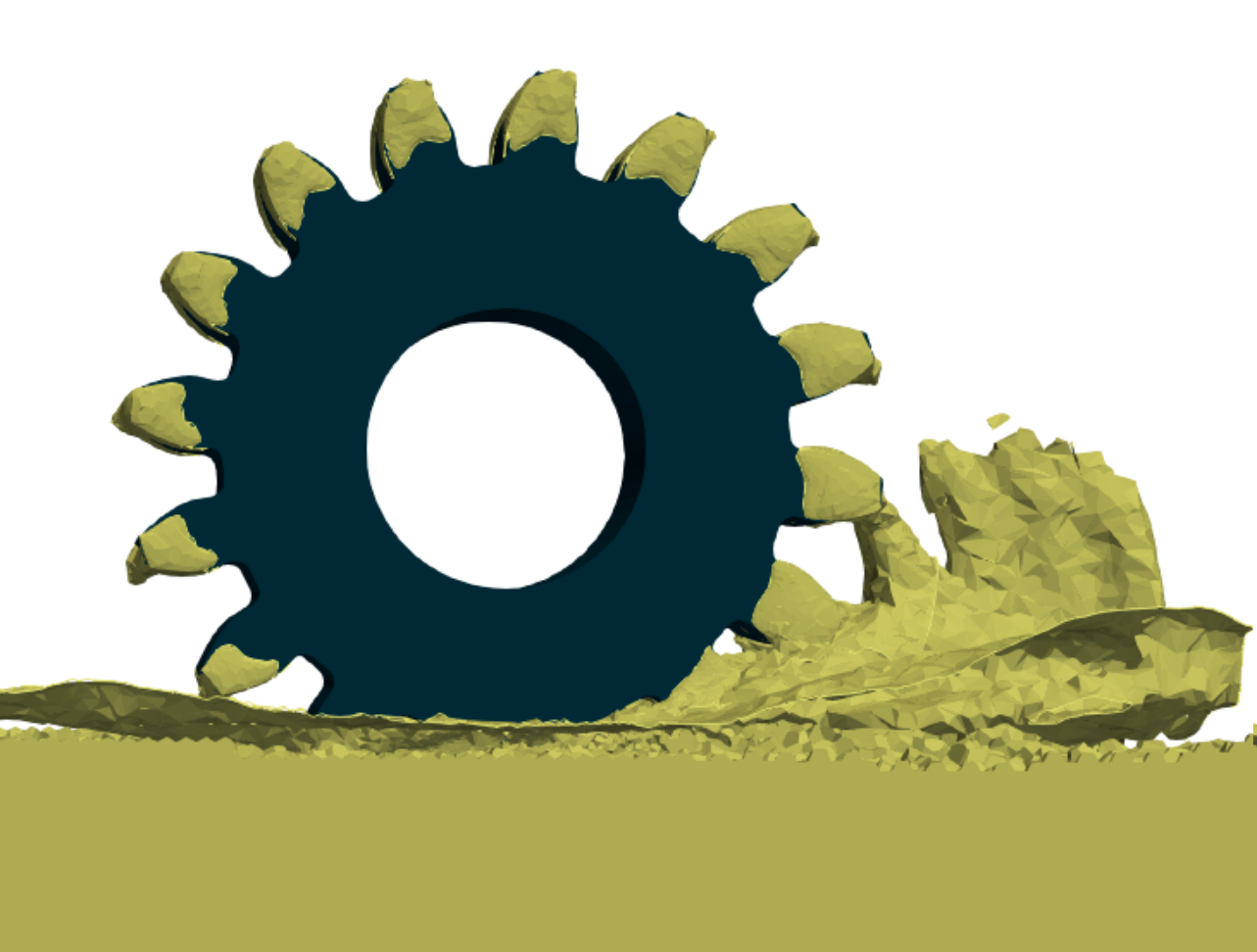}
		\caption{Medium resolution\newline\hspace*{1.5em}Time taken = 124.56 hours}
		\label{fig:OF_FZG_OversetMediumResolution_Result}
	\end{subfigure}
	\begin{subfigure}{0.32\textwidth}
		\centering
		\includegraphics[width=\textwidth]{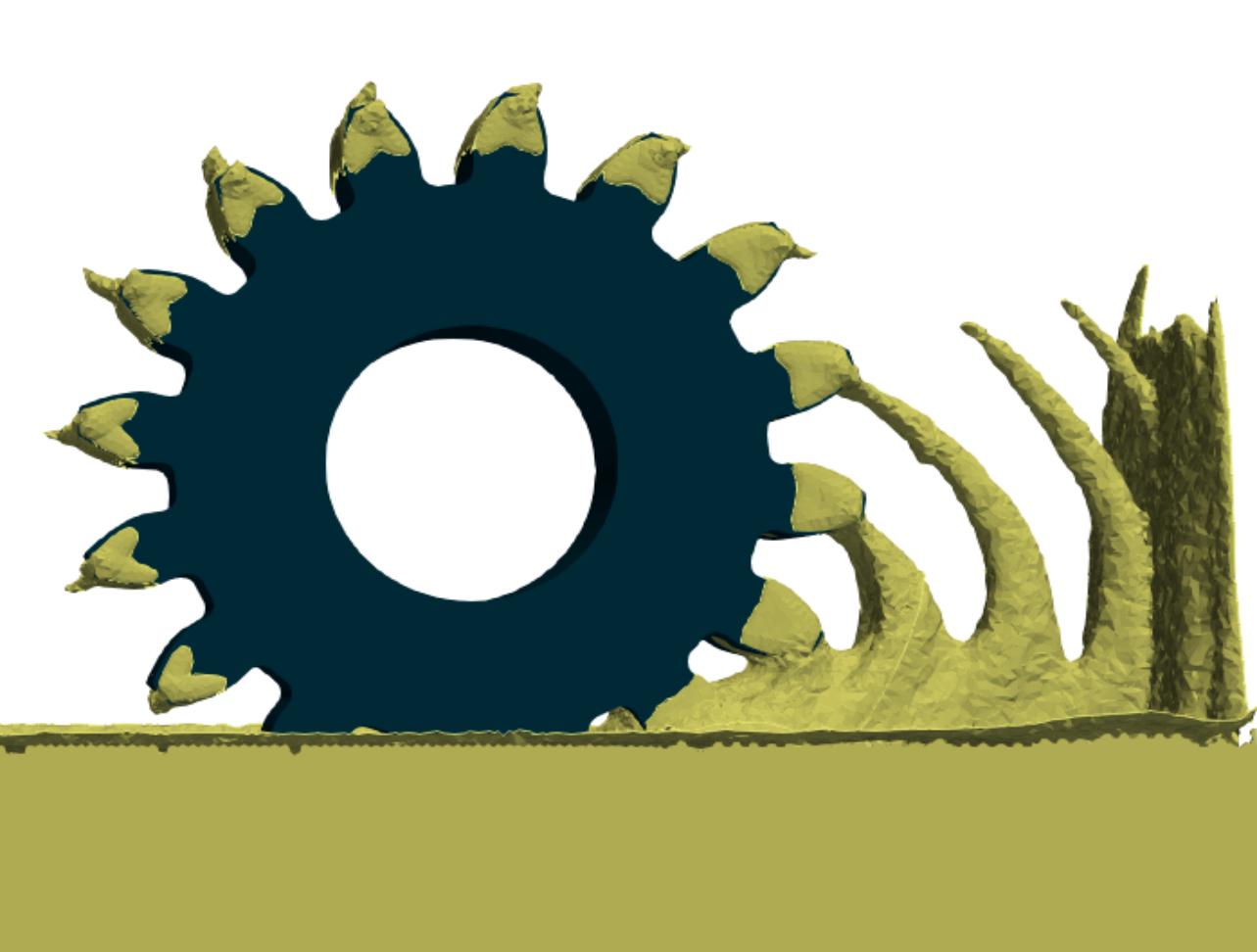}
		\caption{Fine resolution\newline\hspace*{1.5em}Time taken = 273.98 hours}
		\label{fig:OF_FZG_OversetFineResolution_Result}
	\end{subfigure}
    \\
	\caption{Comparison of fluid distribution at different mesh resolution using \textbf{Ansys-Fluent} with overset approach. }
	\label{fig:FZG_AnsysFluent_MeshStudy}
\end{figure}

For the overset mesh approach, the same three resolutions (4, 7, and 10 cells) are evaluated, see \cref{fig:FZG_AnsysFluent_MeshStudy}. To ensure mass conservation, the air phase is modelled using an equation of state. At 4- and 7-cell resolutions, the oil distribution is found to be under-resolved, lacking sufficient detail in fluid structures. The 10-cell configuration, however, provides a noticeably improved representation of the oil distribution. Further mesh refinement was deemed impractical due to the excessive computation cost - approximately 20 days for 0.1 seconds of physical time using the same number of CPUs as in the previous studies. Therefore, the 10-cell configuration is chosen for further comparison using the overset approach.

\newpage
\textbf{PreonLab}
\begin{figure}[bp!]
	\centering
	\includegraphics[width=\textwidth]{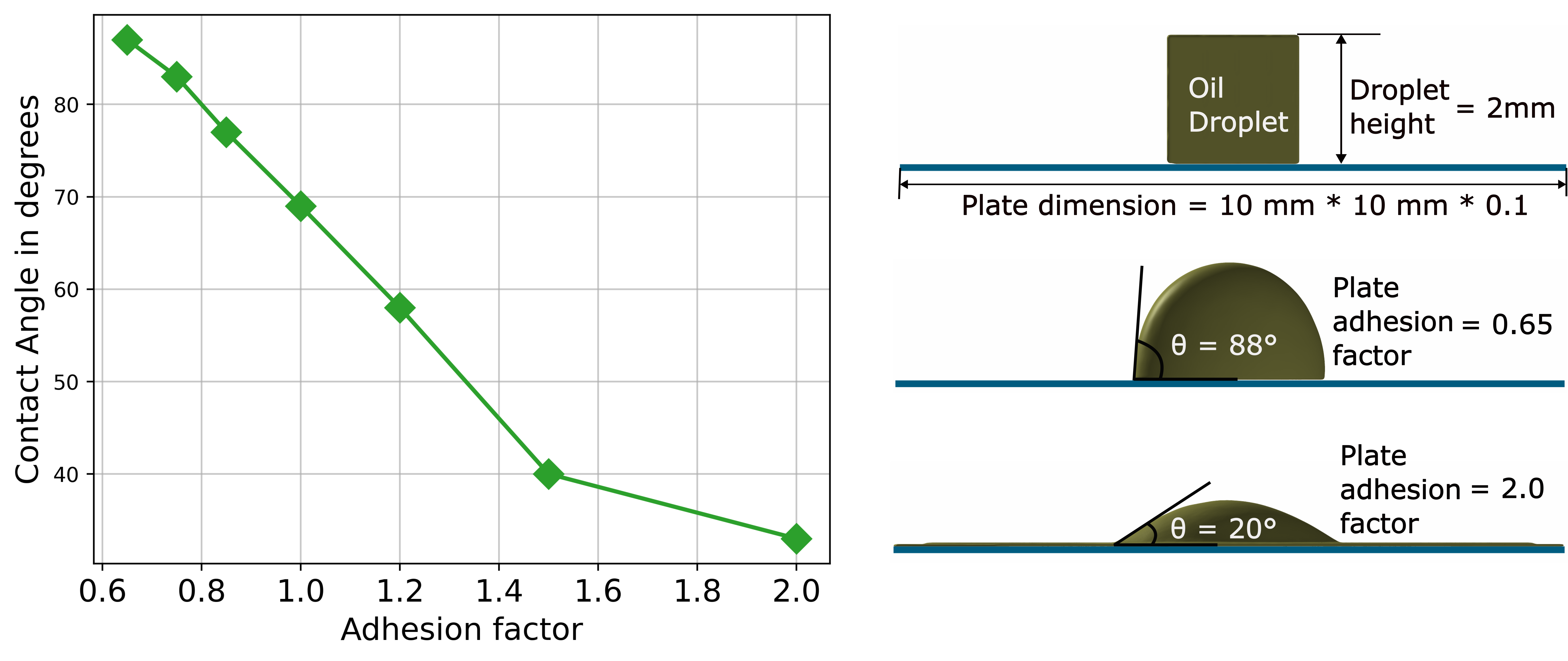}
	\caption{Influence of adhesion factor on contact angle and droplet shape in PreonLab. Left: Plot showing the contact angle for different adhesion factor. Right: The top picture shows the initial position of the droplet and the pictures below show the deformation of droplet for two different adhesion factors (\(\theta\) is the contact angle).}
    \label{fig:PL_Droplet}
\end{figure}

Before conducting the oil distribution simulations in PreonLab, the surface tension model is evaluated to determine appropriate adhesion parameter. PreonLab models surface tension using the potential force approach, where the contact angle between the liquid and solid phases is governed by the balance between adhesion and cohesion forces.

To identify a suitable adhesion value, a preliminary simulation is performed in which a small oil droplet---having the same properties as listed in \cref{tab:LubricantParameter}---is placed on a flat plate. The plate adhesion parameter is systematically varied and the resulting contact angle between the oil and the surface is measured, as illustrated in \cref{fig:PL_Droplet}. Based on these results, an adhesion value of 1.4 corresponding to a contact angle of approximately 45° is selected. This parameter is used consistently in all subsequent PreonLab simulations.

\begin{figure}[tp!]
	\centering
	\begin{subfigure}{0.30\textwidth}
		\centering
		\includegraphics[width=\textwidth]{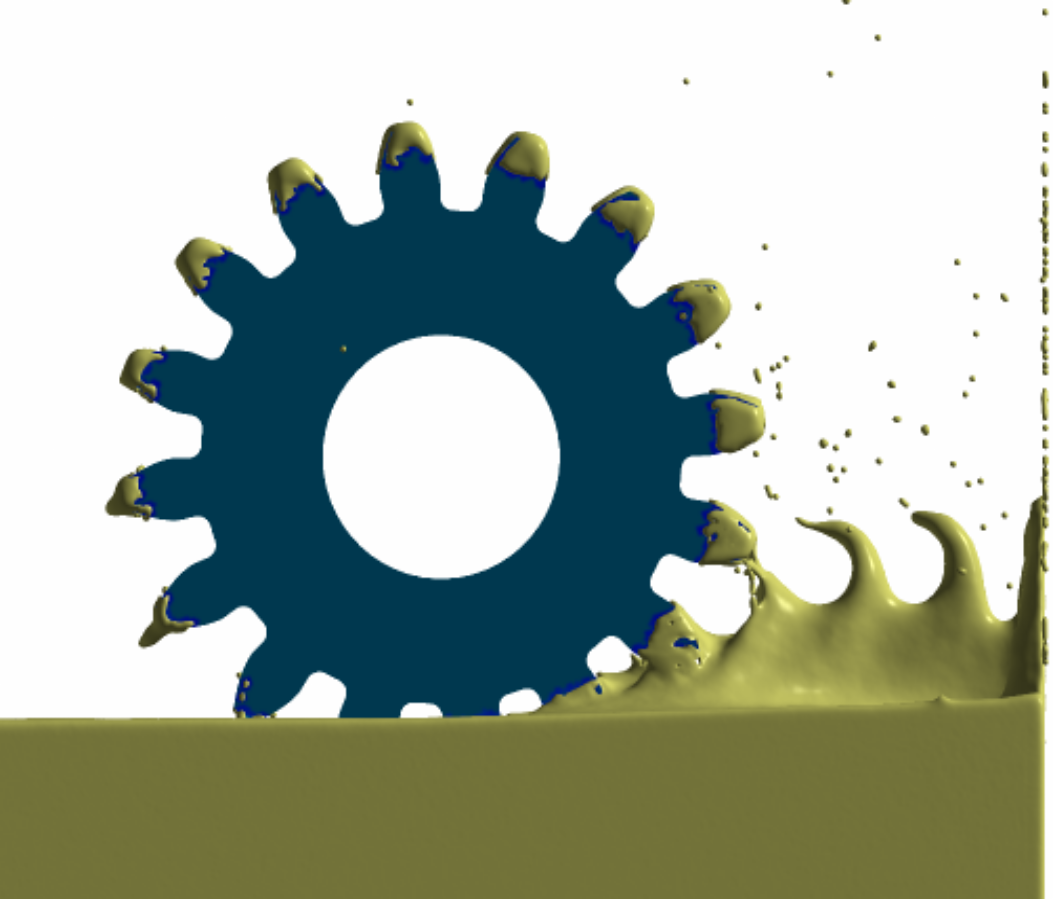}
		\caption{Coarse resolution\newline\hspace*{1.5em}Time taken = 0.38 hours}
		\label{fig:PL_Coarse_ParticleResolution_Result}
	\end{subfigure}
	\begin{subfigure}{0.30\textwidth}
		\centering
		\includegraphics[width=\textwidth]{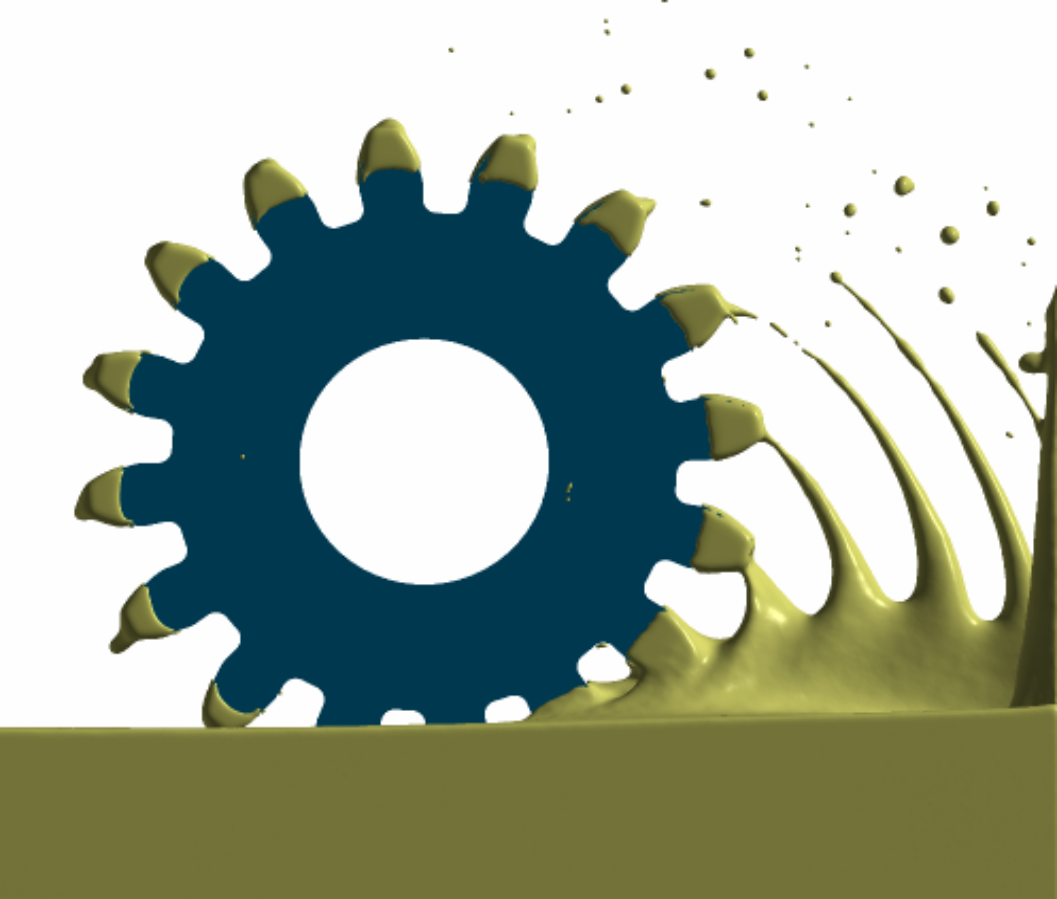}
		\caption{Medium resolution\newline\hspace*{1.5em}Time taken = 2.33 hours}
		\label{fig:PL_Medium_ParticleResolution_Result}
	\end{subfigure}
	\begin{subfigure}{0.30\textwidth}
		\centering
		\includegraphics[width=\textwidth]{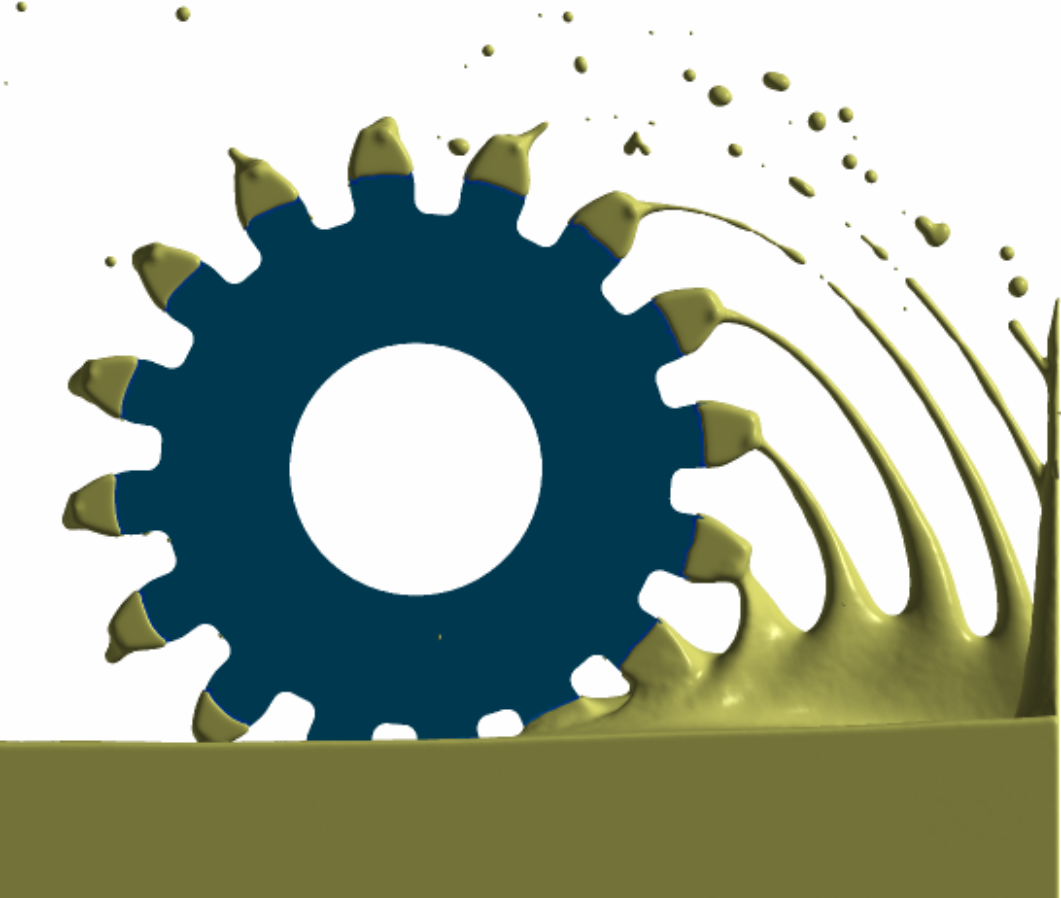}
		\caption{Fine resolution\newline\hspace*{1.5em}Time taken = 6.63 hours}
		\label{fig:PL_Fine_ParticleResolution_Result}
	\end{subfigure}
    \\
	\caption{Comparison of fluid distribution at different particle resolutions with \textbf{PreonLab}. }	\label{fig:PL_FZG_ResolutionResults}
\end{figure}

Following the approach in previous sections, a particle resolution study is conducted in PreonLab using three configurations: 4, 7, and 10 particles between the gear teeth. \Cref{fig:PL_FZG_ResolutionResults} presents the oil distribution for each particle resolution. The distribution varies significantly between 4 and 7 particle configurations. In the 4-particle setup, oil fingers detach from the gear teeth and are visible only on two teeth above the oil sump. In contrast, the 7 particle configuration shows oil fingers extending upto four gear teeth above the sump. While differences between the 7 and 10 particle configurations are minor, the 10-particle configuration displays oil fingers up to five teeth above the sump. Therefore, the 10-particle configuration is selected for further analysis. 

\textbf{MESHFREE}
\begin{figure}[b!]
	\centering
	\begin{subfigure}{0.3\textwidth}
		\centering
		\includegraphics[width=\textwidth]{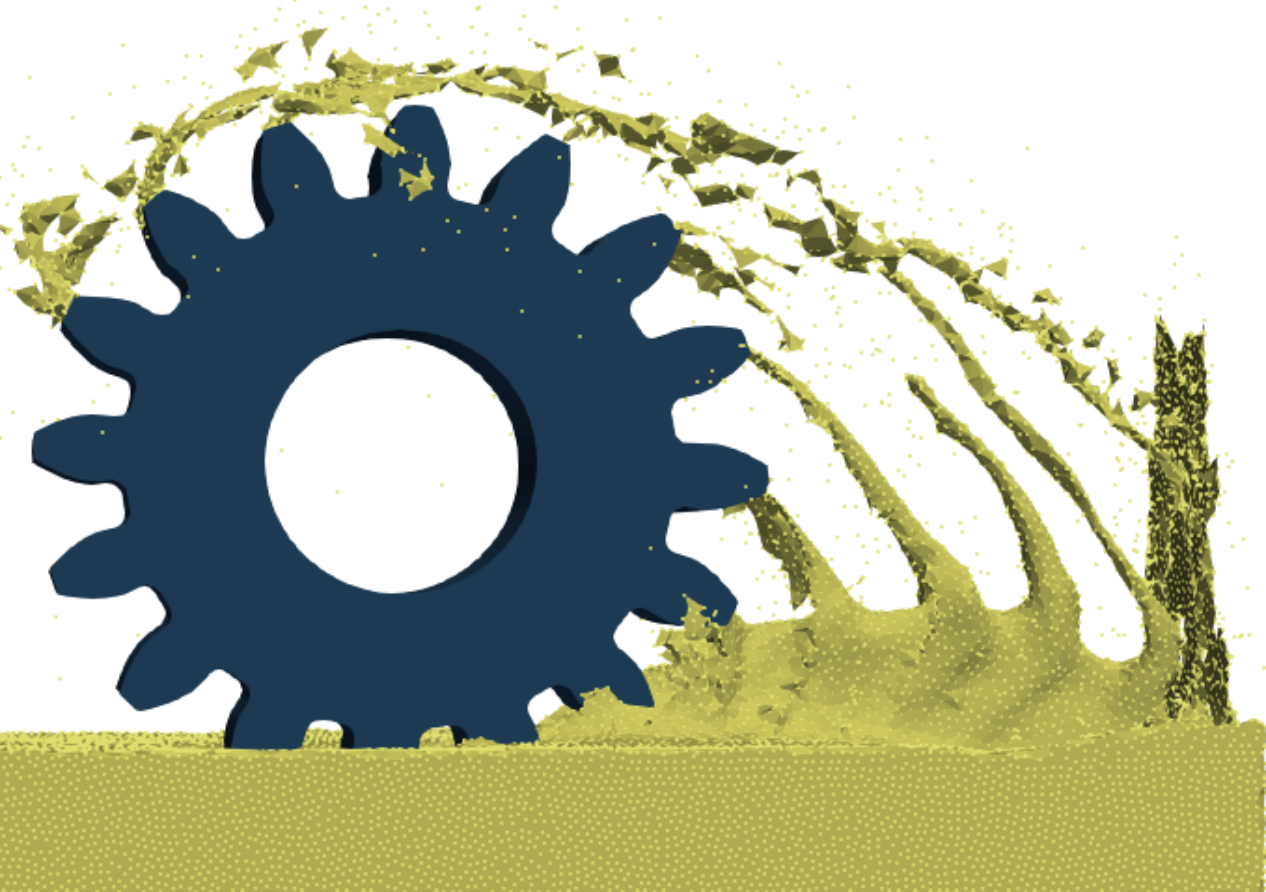}
		\caption{Coarse resolution\newline\hspace*{1.5em}Time taken = 0.81 hours}
		\label{fig:MF_FZG_CoarseMesh_Results}
	\end{subfigure}
	\begin{subfigure}{0.3\textwidth}
		\centering
		\includegraphics[width=\textwidth]{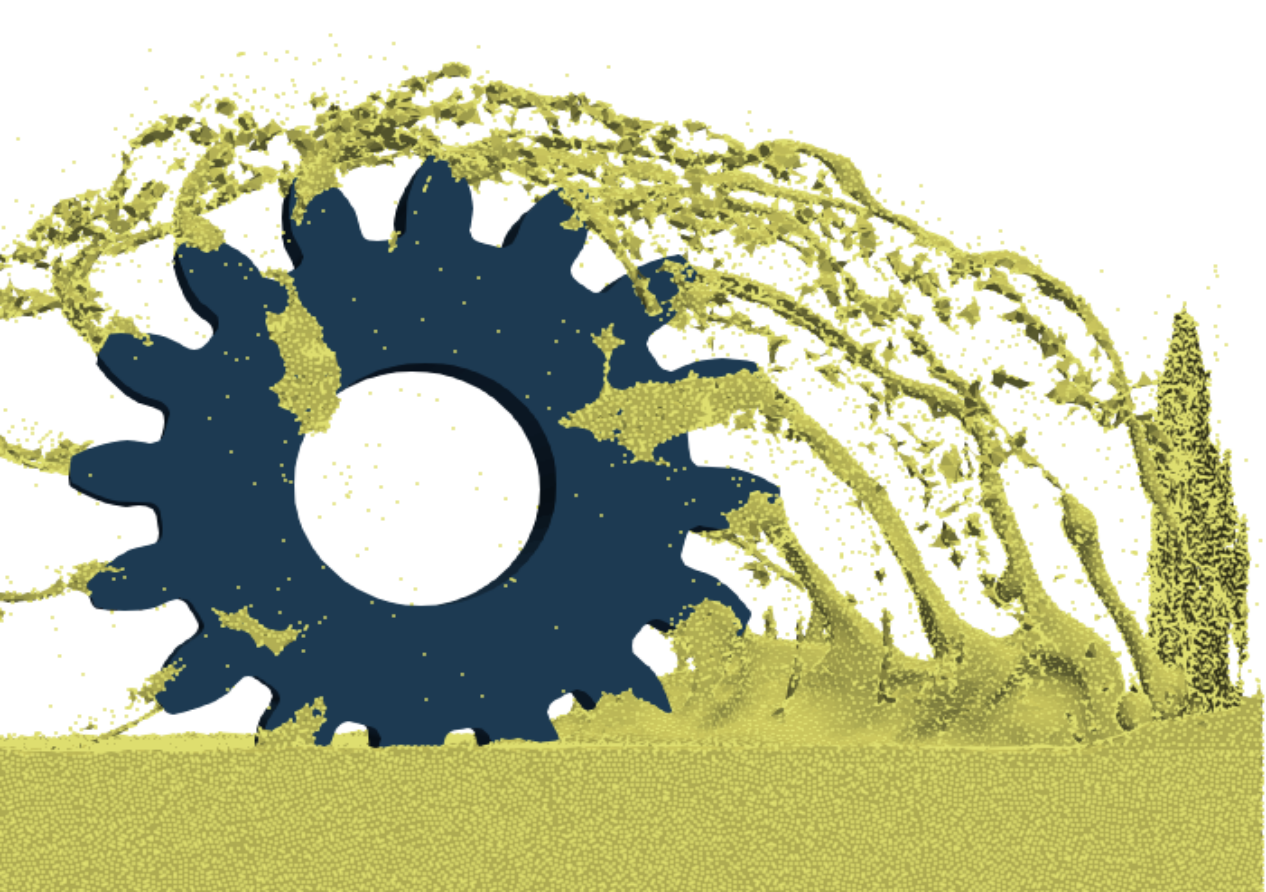}
		\caption{Medium resolution\newline\hspace*{1.5em}Time taken = 6.31 hours}
		\label{fig:MF_FZG_MediumMesh_Result}
	\end{subfigure}
	\begin{subfigure}{0.3\textwidth}
		\centering
		\includegraphics[width=\textwidth]{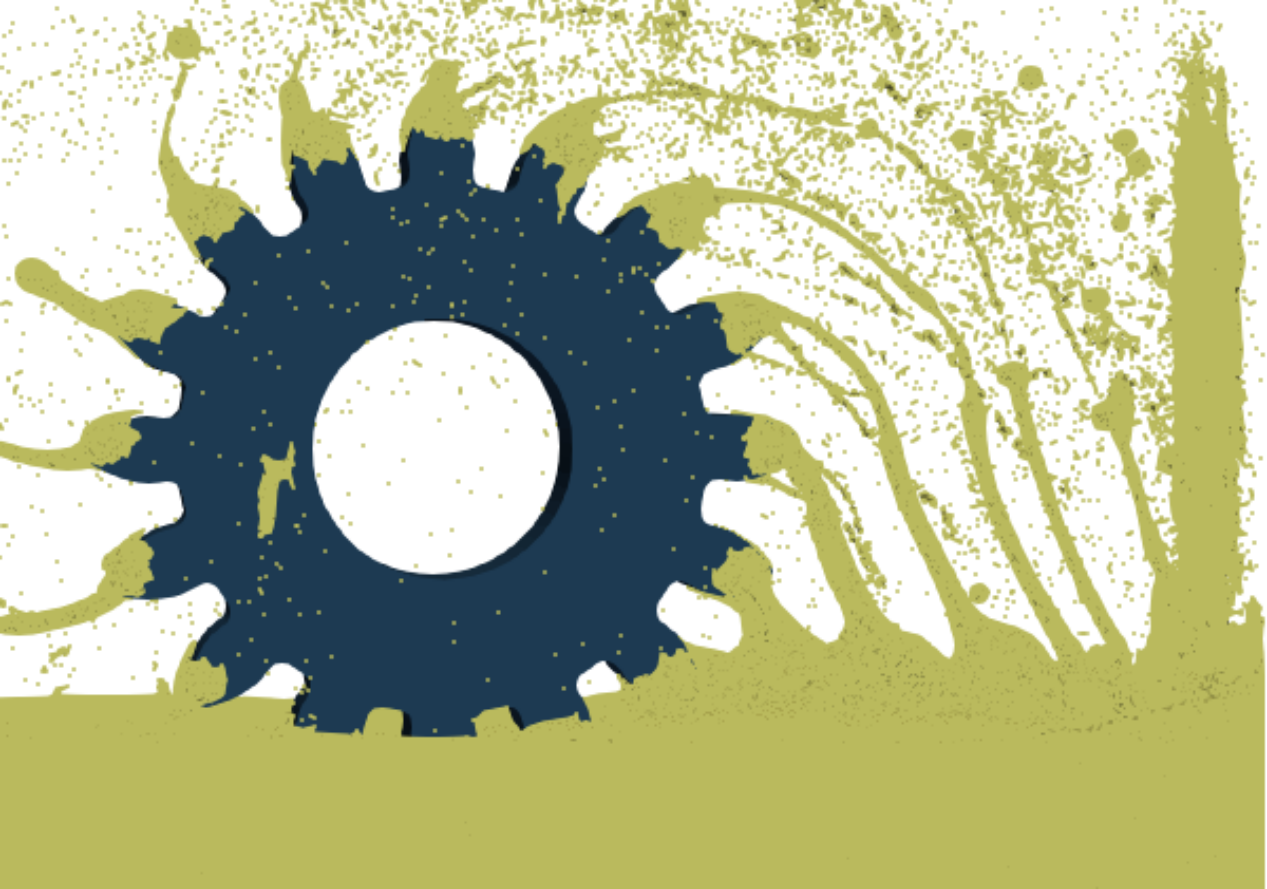}
		\caption{Fine resolution\newline\hspace*{1.5em}Time taken = 72.0 hours}
		\label{fig:MF_FZG_FineMesh_Result}
	\end{subfigure}
    \\
	\caption{Comparison of fluid distribution at different particle resolutions with \textbf{MESHFREE}. }
	\label{fig:MF_FZG_Result}
\end{figure}
In the MESHFREE simulation, a particle resolution analysis is conducted using three resolutions: 8, 12, and 24 particles between consecutive gear teeth as resolutions below 8 particles are considered under-resolved. \Cref{fig:MF_FZG_Result} presents the oil distribution for each particle resolution. At the coarse resolution of 8 particles, oil fingers are seen up to the third gear tooth above the oil sump. In the 12-particle configuration, oil fingers are visible on all teeth, with additional oil deposition appearing on the top of the gear adjacent to the fourth tooth above the sump. However, the distribution of oil remains rather erratic. The 24-particle configuration similarly displays oil fingers on all teeth, but with a notably smoother distribution pattern compared to the lower resolutions. Given the improved consistency in oil distribution, the 24-particle configuration is selected for further analysis.

\newpage
\textbf{Comparison of solvers}   

\begin{figure}[htbp!]
	\centering
	\includegraphics[width=\textwidth]{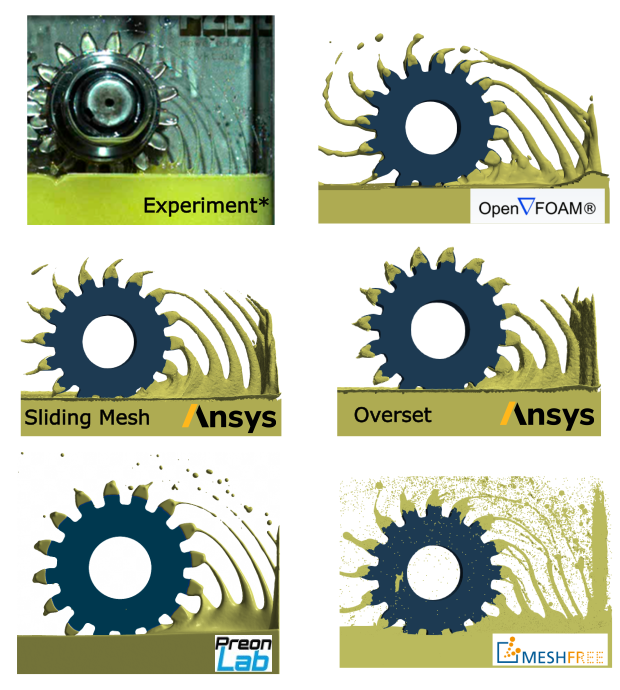}
	\caption{Qualitative results and comparison of different solvers with experimental results for oil distribution in a FZG gearbox.}
	\label{fig:FZG_AllSoftware_Visualization}
\end{figure}

This section compares the optimal configurations identified for each solver. Configuration details are summarized in \cref{tab:FZG_ComparisonConfiguration}, with visualizations for each software shown in \cref{fig:FZG_AllSoftware_Visualization}, and the corresponding simulation times presented in \cref{fig:FZG_AllSoftware}. To ensure consistency, all simulations are run for a physical time of one second, using identical hardware and the same number of CPU cores. 
\begin{table}[b]
    \begin{tabular}{ p{2.5cm}p{2.5cm}p{2.5cm}p{2.5cm}p{2.5cm}}
     \toprule
     Solver & Method & Discretization & Mesh-motion & Resolution\\
    \midrule
     interFoam & algebraic VoF &unstructured hexahedral &  AMI/sliding & 10\\
     Fluent & algebraic VoF & unstructured polyhedral &  AMI/sliding overset & 10 \\  
     PreonLab & SPH & only liquid& & 10\\
     Meshfree & FPM & only liquid & & 24\\
    \bottomrule
    \end{tabular}
    \caption{Resolution selected for qualitative comparison in each solver.}
    \label{tab:FZG_ComparisonConfiguration}
\end{table}
Across all solvers, the simulated oil distribution closely matches experimental observations, with PreonLab demonstrating the highest accuracy. Addition to its superior fidelity, PreonLab also completes the simulation in the shortest time. 

\begin{figure}[tb!]
	\centering
	\begin{subfigure}{0.32\textwidth}
		\centering
		\includegraphics[width=\textwidth]{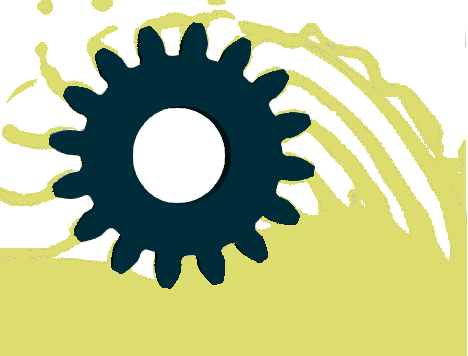}
        \caption{OpenFOAM with Arbitrary Mesh Interface.}
		\label{fig:FZG_OpenFOAM_AlphaField}
	\end{subfigure}
    \begin{subfigure}{0.32\textwidth}
    	\centering
    	\includegraphics[width=\textwidth]{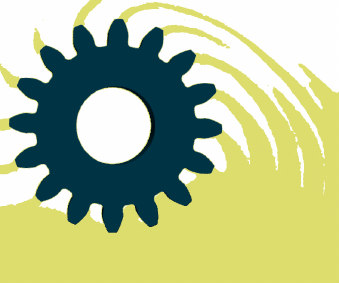}
        \caption{Ansys-Fluent with sliding mesh approach.}
    	\label{fig:FZG_Ansys_SM_AlphaFieldValue}
    \end{subfigure}
    \begin{subfigure}{0.32\textwidth}
    	\centering
    	\includegraphics[width=\textwidth]{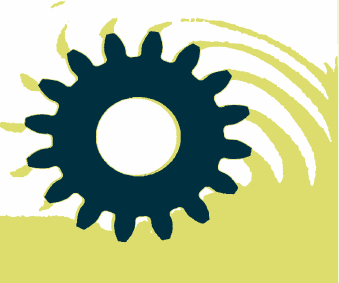}
        \caption{Ansys-Fluent with overset approach.}
    	\label{fig:FZG_Ansys_Overset_AlphaField}
    \end{subfigure}
	\caption{Alpha field visualization of oil distribution with mesh based methods. }	\label{fig:FZG_AlphaField}
\end{figure}

While simulations using OpenFOAM and Ansys-Fluent with the AMI or sliding mesh approach require longer computation times, the resulting oil distribution remains in good agreement with experimental observations. The discrepancies seen are attributed to the post-processing methodology in ParaView. The visualization of the alpha field, as shown in \cref{fig:FZG_AlphaField}, confirms that the oil distribution closely matches the experimental results. In both OpenFOAM and Ansys-Fluent sliding mesh simulations, oil fingers extend up to the fifth gear tooth above the oil sump, and an oil lamella is observed along the housing wall. Despite the good agreement in results, the high computational cost of this approach limits its practicality. Furthermore, the AMI/sliding mesh method presents significant limitations when simulating two rotating gears. Specifically, when the interpolation interface regions overlap, leading to numerical instability and failure of the simulation.
The simulation using Ansys-Fluent with the overset mesh approach takes the longest - approximately three times longer than the sliding mesh approach, and nearly ten times longer than PreonLab. In addition to the high computational cost, this setup shows the oil finger detaching at the fourth gear tooth, with no substantial accuracy gain over the sliding mesh method.

The simulation time for MESHFREE lies between that of PreonLab and the mesh-based solvers, however, it requires a higher resolution. While the overall oil distribution around the gear teeth appears comparable with the experiments, the results show dispersed particles within the housing volume. This leads to an irregular oil pattern compared to the other solvers. Furthermore, oil fingers are predicted on almost every gear tooth, indicating a noticeable overestimation of this flow structure that is not supported by the experimental findings.

\subsection{Quantitative comparison}
\label{sec:Mauz_Quantitative}

The experimental setup and results presented in the thesis of Mauz~\cite{mauz-hvvsbub6m1988} serve as the basis for the quantitative validation and comparison of the four simulation tools assessed in this study. 
\newfloatcommand{capbtabbox}{table}[][\FBwidth]
\begin{figure}[t]
    \begin{floatrow}
    \ffigbox{%
               \centering
              \graphicspath{{Figures/}}
              \def\svgwidth{0.5\textwidth}
              \normalsize
              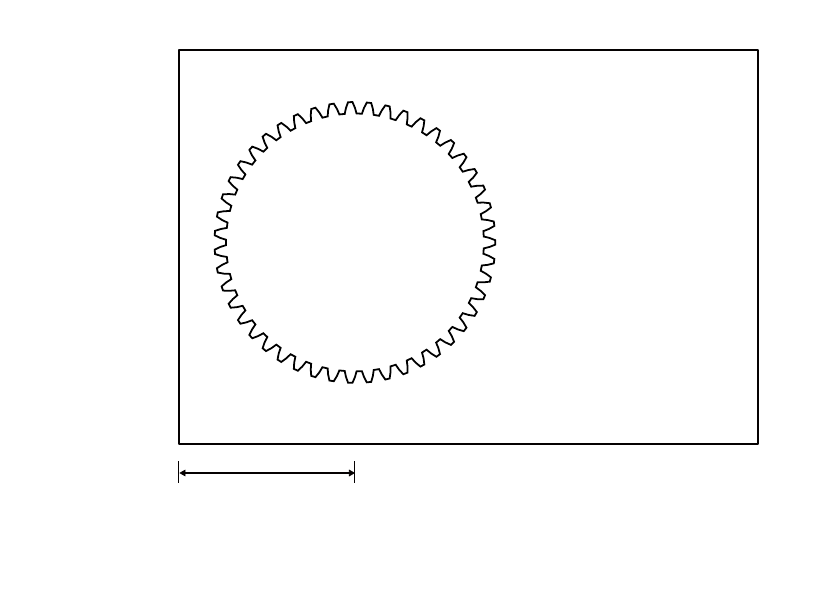
    }{%
      \caption{Schematic diagram showing the numerical setup used for quantitative validation.}%
      \label{fig:MauzNumericalSetup}
    }
    \capbtabbox{%
            \begin{tabular}{ p{4.5cm}p{1.2cm}}
                 \toprule
                 Parameter & Value \\
                \midrule
                Number of teeth                 & 47            \\
                Module in mm       & 4.5            \\
                Pressure angle in degrees              & 20            \\
                Pitch diameter $R$ in mm              & 105.75         \\
                \bottomrule
            \end{tabular}
    }{%
      \caption{Geometric properties of the gear used for quantitative validation.}%
      \label{tab:Mauz_GearProperty}
    }
    \end{floatrow}
\end{figure}

The numerical configuration used for the simulations is depicted in \cref{fig:MauzNumericalSetup} and features a single gear, the properties of which are listed in \cref{tab:Mauz_GearProperty}. The gear rotates at a speed of 868 RPM within a lubricant-filled housing with dimensions $L = $ 480 mm (length), $H = $ 300 mm (height) and $W = $ 110 mm (width). The gear is positioned at a distance $X_c =$ 150 mm from the left wall and $Y_c =$ 150 mm from the bottom wall of the housing. The level of lubricant fill, denoted by $e$, corresponds to a vertical height of 55 mm measured from the tip of the gear to the surface of the lubricant.

The validation study is conducted using three different lubricants, with their properties summarized in \cref{tab:Mauz_OilProperties}. For each solver, a mesh resolution comparable to that described in the above section is employed. To ensure consistency across all simulations, the surface tension and contact angle are held constant at 0.032 N/m and 45°, respectively. For the quantitative comparison, the overset approach in Ansys-Fluent is excluded due to its significantly high computational cost, as observed in the previous analysis.

\begin{table}[b]
	\centering
	\caption{Properties of the lubricant used for quantitative analysis.}
	\label{tab:Mauz_OilProperties}
	\begin{tabular}{@{}lccc@{}}
		\toprule
		Parameter & Lubricant 1 & Lubricant 2 & Lubricant 3 \\
		\midrule
		Density in $kg/m^3$                   & 842  & 855 & 881     \\
		Kinematic viscosity in $mm^2/s$       & 14   & 30  & 240     \\
		\bottomrule
	\end{tabular}
\end{table}

\begin{figure}[b!]
	\centering
	\includegraphics[width=0.99\textwidth]{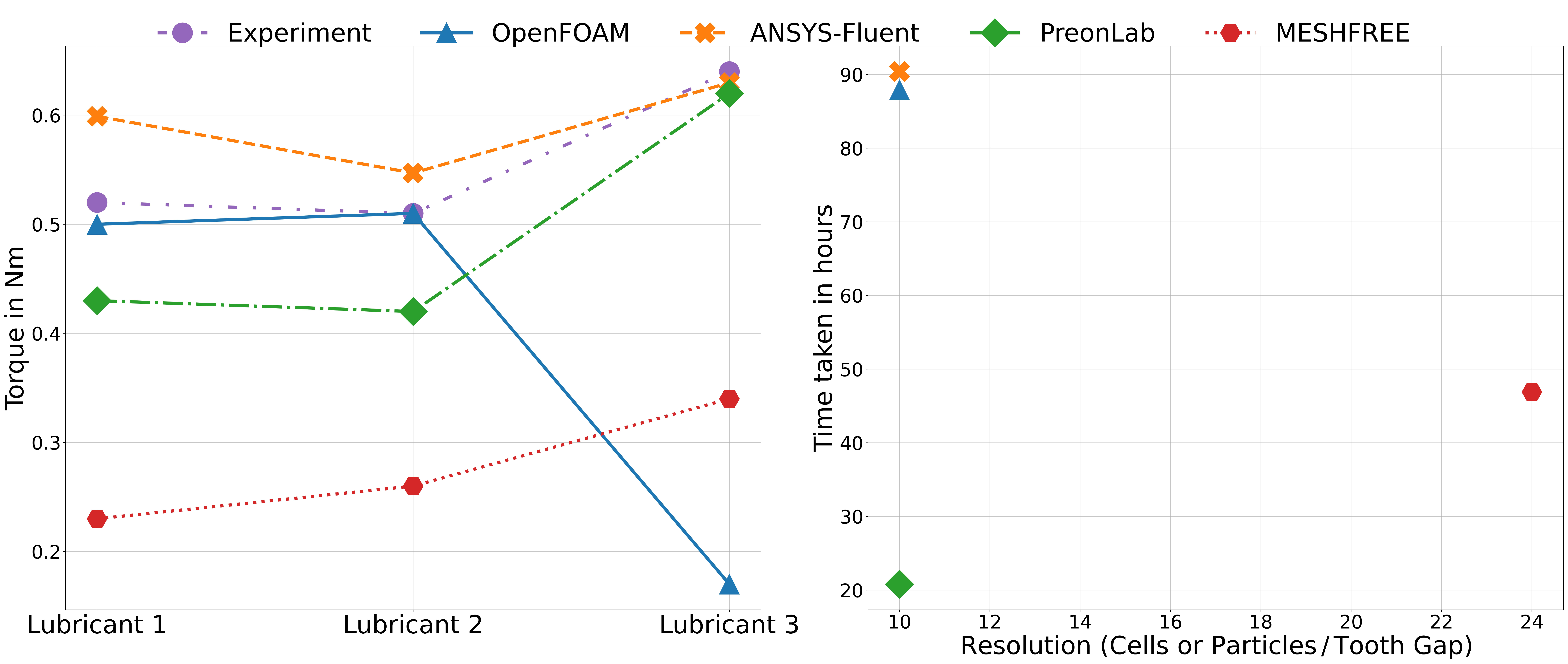}
	\caption{ Plots showing the results of quantitative study. Left: Plot showing the torque acting on the rotating gear for three lubricants: Experiments vs Simulations. Right: Resolution and simulation time compared for lubricant 2 for each solver at optimum configuration.}
	\label{fig:Mauz_TimeTaken_AllSoftware}
\end{figure}

\Cref{fig:Mauz_TimeTaken_AllSoftware} presents a comparison between the torque acting on the gear as predicted by simulations and the corresponding experimental measurements. For lubricants 1 and 2, OpenFOAM exhibits the lowest deviation among the tested solvers. However, for lubricant 3, which has significantly higher viscosity, the deviation increases sharply to approximately 70\%. Further simulations with higher viscosities confirms that OpenFOAM struggles with highly viscous fluids. Ansys-Fluent shows good agreement with experimental results across all lubricants, with deviations ranging from around 15\% for lubricant 1 to as low as 1\% for lubricant 3. However, as noted in the previous section, the AMI/sliding mesh approach used in OpenFOAM and Ansys-Fluent is not suitable for configurations involving more than one rotating gear, due to overlapping interface regions that lead to simulation failure. PreonLab demonstrates relatively consistent performance, with deviations of approximately 17\% for lubricants 1 and 2. In contrast to OpenFOAM, PreonLab performs well with the high-viscosity lubricant 3, achieving a deviation as low as 3\%. MESHFREE, on the other hand, shows a consistent deviation of approximately 50\% across all lubricants, indicating an issue with the interaction between gear and acting fluid.

In terms of computational cost, PreonLab offers the best performance, completing the simulation in approximately 20 hours for the selected resolution. To ensure consistency, all simulations are executed on the same hardware and cover a physical time duration of one second. The simulation times for lubricant 2 are presented in \cref{fig:Mauz_TimeTaken_AllSoftware}. While OpenFOAM achieves the lowest torque deviation for lubricants 1 and 2, it incurs a significantly higher computational cost---approximately four times that of PreonLab. Ansys-Fluent exhibits a comparable computational cost to OpenFOAM. MESHFREE falls between PreonLab and the mesh-based solvers in terms of simulation time, but shows a higher deviation from the experimental torque.

\section{Grease distribution in gearbox}
Based on the previous results, only PreonLab is selected for the subsequent study on grease distribution. 

For the simulation of non-Newtonian fluids, the Herschel-Bulkley model in PreonLab is used as 
\begin{equation}
    \eta = \frac{\tau_{HB}}{\dot{\gamma}} (1 - e^{-m\dot{\gamma}}) + c \dot{\gamma}^{n-1}
\end{equation}
where \(\tau_{HB}\) is the yield stress, \(c\) is the flow consistency index, and n is the flow behaviour index, \(m=M \frac{L}{V}\) is the stress growth exponent factor, as described in \cite{SPH_ST2}. Here, M is initialized with a value greater than 50 and progressively increased to balance accuracy and computational cost, \(L\) is a reference length, and \(V\) is a reference velocity.

In this section, the grease distribution in a gearbox is first validated using  experimental results from Liu et al. \cite{LiuGrease}, considering two filling levels: 40\% and 54\% of the housing volume. This is followed by a case study investigating the influence of gear speed and filling volume on the amount of grease deposited around the gear and the pinion.

\subsection{Qualitative Comparison}

\begin{figure}[tbp!]
	\centering
	\includegraphics[width=0.7\textwidth]{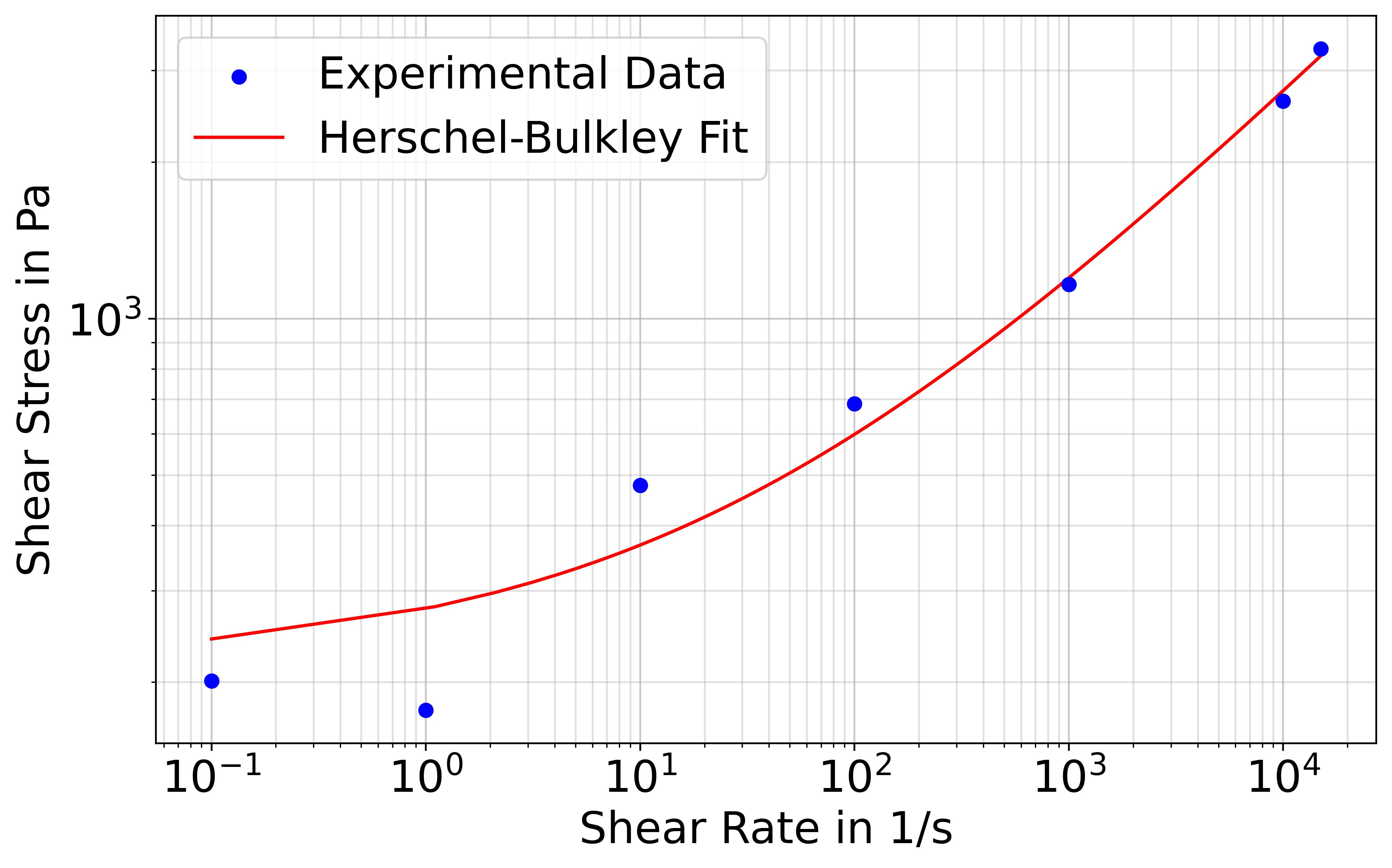}
	\caption{Plot showing the curve fitting for Herschel-Bulkley model with experimental data.}
	\label{fig:GreaseCurveFitting}
\end{figure}

\begin{figure}[h!]
    \centering
    \begin{subfigure}{0.99\textwidth}
        \includegraphics[width=\textwidth]{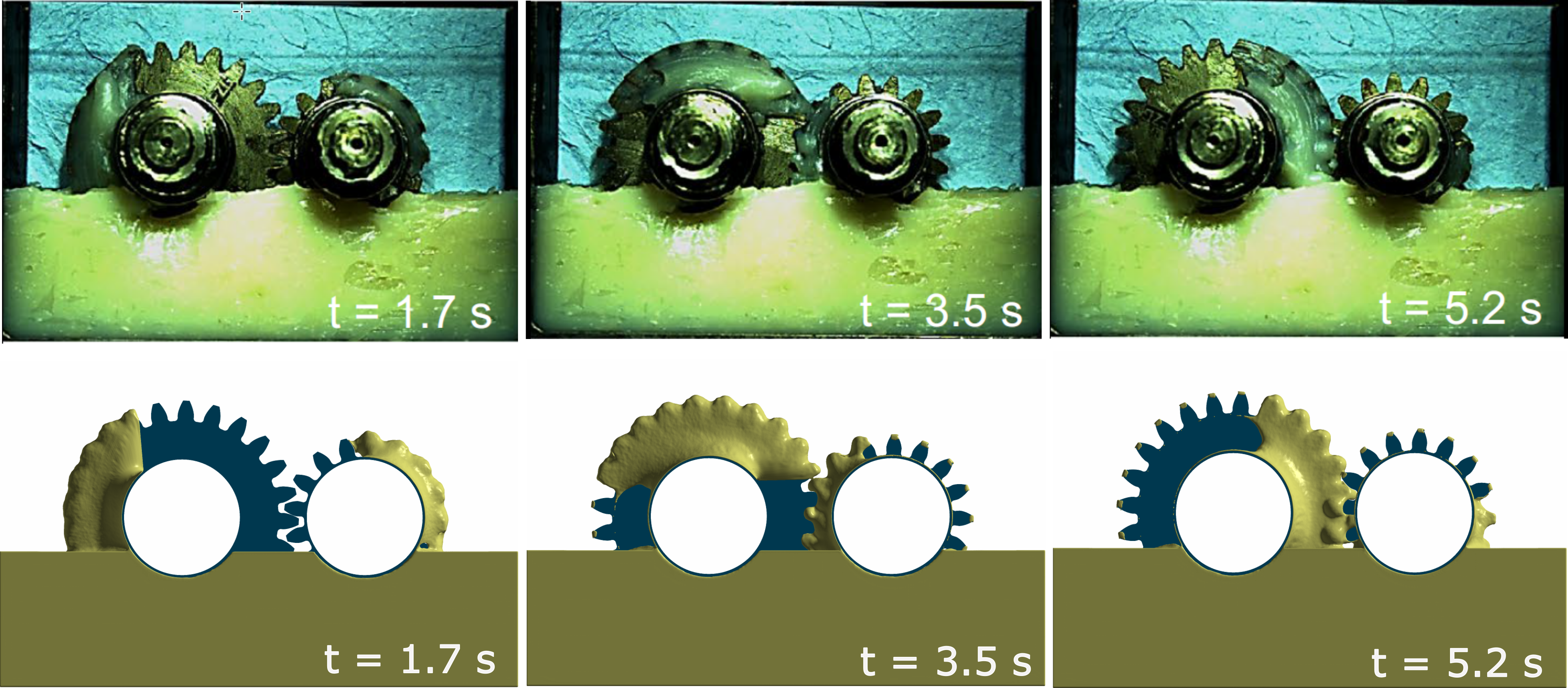}
        \caption{40\% grease filling. Top: Experimental results from \cite{LiuGrease}. Bottom: PreonLab simulation results.}
        \label{fig:FZG_GreaseDistribution_40Percent}
    \end{subfigure}
    
    \vspace{0.5cm} 

    \begin{subfigure}{0.99\textwidth}
        \includegraphics[width=\textwidth]{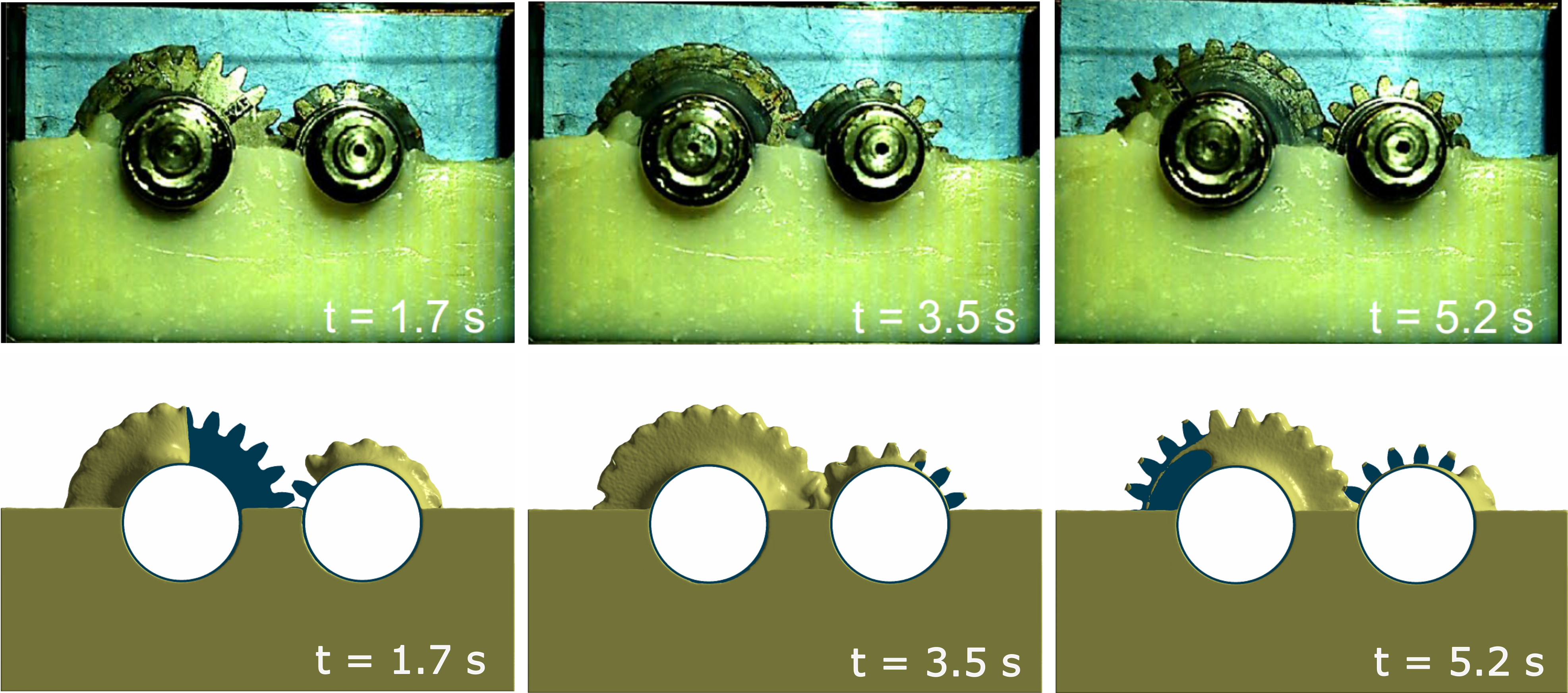}
        \caption{54\% grease filling. Top: Experimental results from \cite{LiuGrease}. Bottom: PreonLab simulation results.}
        \label{fig:FZG_GreaseDistribution_54Percent}
    \end{subfigure}

    \caption{Qualitative comparison of grease distribution in an FZG gearbox for different grease filling levels.}
    \label{fig:FZG_GreaseDistribution_Comparison}
\end{figure}

\begin{table}[bp!]
    \RawFloats
    \centering
    \begin{minipage}[t]{0.4\textwidth}
        \vspace{0pt}  
       \centering
	\begin{tabular}{@{}lrr@{}}
		\toprule
		Shear rate &  Grease Viscosity \\ \(\dot{\gamma}\) in 1/s &  \(\eta\) in mPas \\ 
		\midrule
		0.1     & 20121700.00   \\
		1       & 176862.50       \\
		10 & 47838.20        \\
		100 & 6864.04     \\
		1000 & 1163.52     \\
        10000 & 262.06  \\
        15000 & 220.27 \\
		\bottomrule
	\end{tabular}
        \caption{Rheological data of NLGI 1-2 grease.}
	\label{tab:FZG_GreaseProperties}
    \end{minipage}
    \hfill
    \begin{minipage}[t]{0.45\textwidth}
        \vspace{0.9\baselineskip}
        \centering
	\begin{tabular}{@{}ll@{}}
		\toprule
		Parameter &  Value \\ 
		\midrule
		Shear viscosity in Pas     & 1.16352       \\
		Bulk viscosity in Pas       & 0.0      \\
		Flow behaviour index & 0.41       \\
		Yield stress in Pa & 220.0  \\
		Stress growth exponent in s & 118.4    \\
		\bottomrule
	\end{tabular}
        \caption{Properties of the NLGI 1-2 grease used in PreonLab}
        \label{tab:GreasePropertiesPreonLab}
    \end{minipage}
\end{table}
The same gearbox as in \cref{sec:singlegearCase1} is employed. However, in this case, the complete gearbox with two gears is considered, see \cref{fig:FZG_GreaseDistribution_40Percent}. An NLGI 1-2 grease characterized by the rheological properties listed in \cref{tab:FZG_GreaseProperties}, is used. The data is fitted to the Herschel-Bulkley model within PreonLab to determine the material properties as input parameters required for the simulation, see \cref{tab:GreasePropertiesPreonLab} and \cref{fig:GreaseCurveFitting}.

In addition to the parameters listed, an adhesion value of 20 and a cohesion value of 0.5 are used. As discussed in \cref{sec:SPH_Theory} and \cref{sec:singlegGearQualitative}, the adhesion and cohesion parameters in the potential force model must be calibrated appropriately. Unlike the case described in \cref{sec:singlegGearQualitative}, no established method exists to directly relate these parameters to a known physical quantity, such as the contact angle, when modelling grease.

Furthermore, replicating the droplet test used for oils—by placing a grease droplet on a flat plate—does not produce meaningful results due to the inherent rheological behaviour of grease, which prevents significant deformation or movement. As a result, the adhesion and cohesion values must be determined empirically for each grease type by comparing simulation results with experimental observations.

The results of the grease distribution obtained with PreonLab and the corresponding experimental observations are shown in  \cref{fig:FZG_GreaseDistribution_40Percent} and  \cref{fig:FZG_GreaseDistribution_54Percent}. All simulations for grease distribution are conducted using the GPU solver on an NVIDIA H100 GPU.

Overall, the grease distribution predicted from PreonLab simulations show good agreement with experimental results. For a filling level of 40\%, the amount of grease deposited on the gears in the simulation qualitatively matches the experimental observations. Grease is initially scooped out of the sump and deposited on the gear and pinion surfaces. This initial scoop creates a void in the grease sump; as a result, in subsequent time steps, no additional grease is deposited on the gear or pinion teeth. At a filling level of 54\%, the grease deposited between the gear and pinion teeth also aligns with the experimental findings. However, the grease on the gear and pinion surfaces appears over-predicted in the simulations. In experiments, the amount of grease on these surfaces is noticeably lower. Only a small line of grease can be seen on the top of shafts. It is important to note here that the grease distribution is highly sensitive to the initial condition in experiments - particularly the surface properties, such as purity and the absence of residual grease or oil from previous runs. Even slight changes in these properties can result in a completely different grease distribution. Despite some deviations from experimental data, PreonLab results for 54\% filling level are considered reliable. Therefore, this setup is further used to study the influence of gear speed and filling volume on grease distribution.



\subsection{Influence of gear speed and filling volume on the amount of grease deposited on the gears}
In this section, the influence of gear speed and filling volume on the amount of grease deposited on the gear and pinion is investigated, due to its high importance in technical applications. 

\begin{figure}[t!]
	\centering
	\begin{subfigure}{0.99\textwidth}
		\centering
		\includegraphics[width=\textwidth]{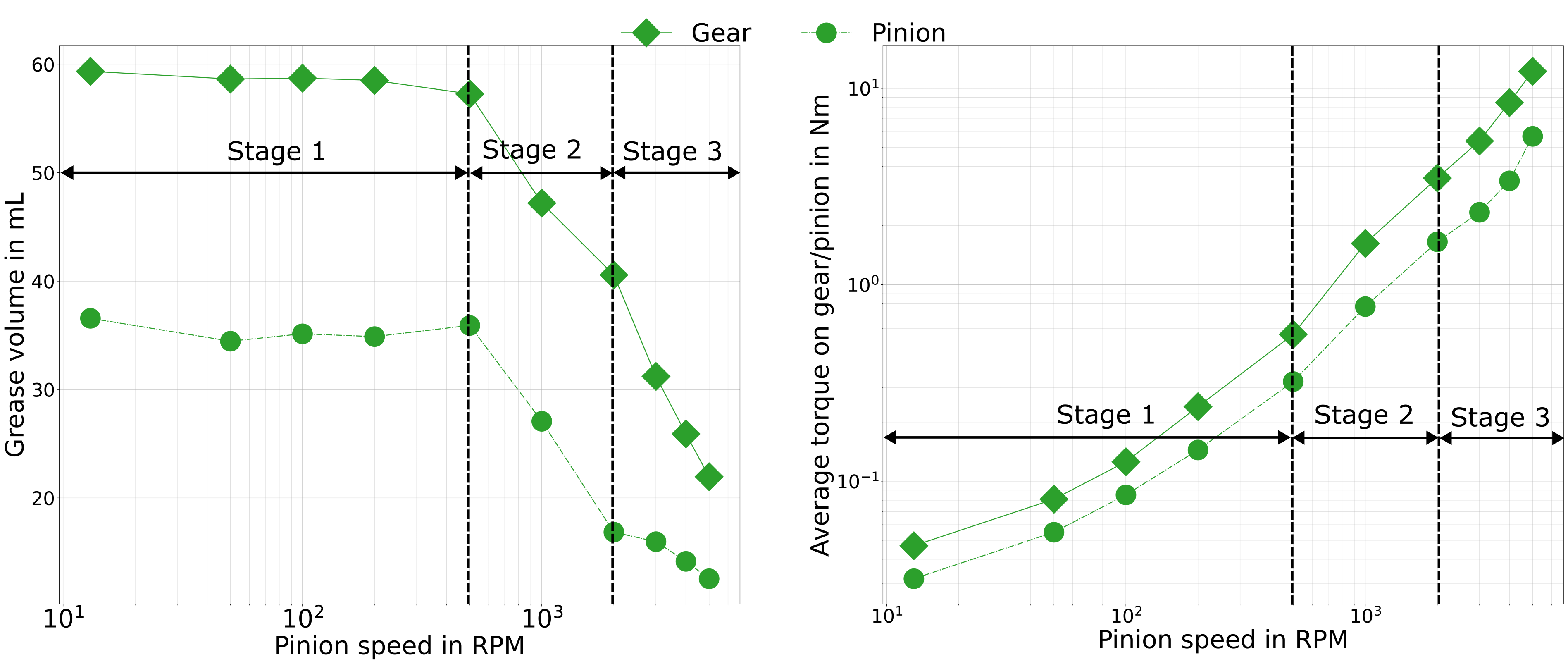}
		\label{fig:FZG_Grease_CaseStudy_GearSpeed_Plot}
	\end{subfigure}
        \\
	\begin{subfigure}{0.43\textwidth}
		\centering
		\includegraphics[width=\textwidth]{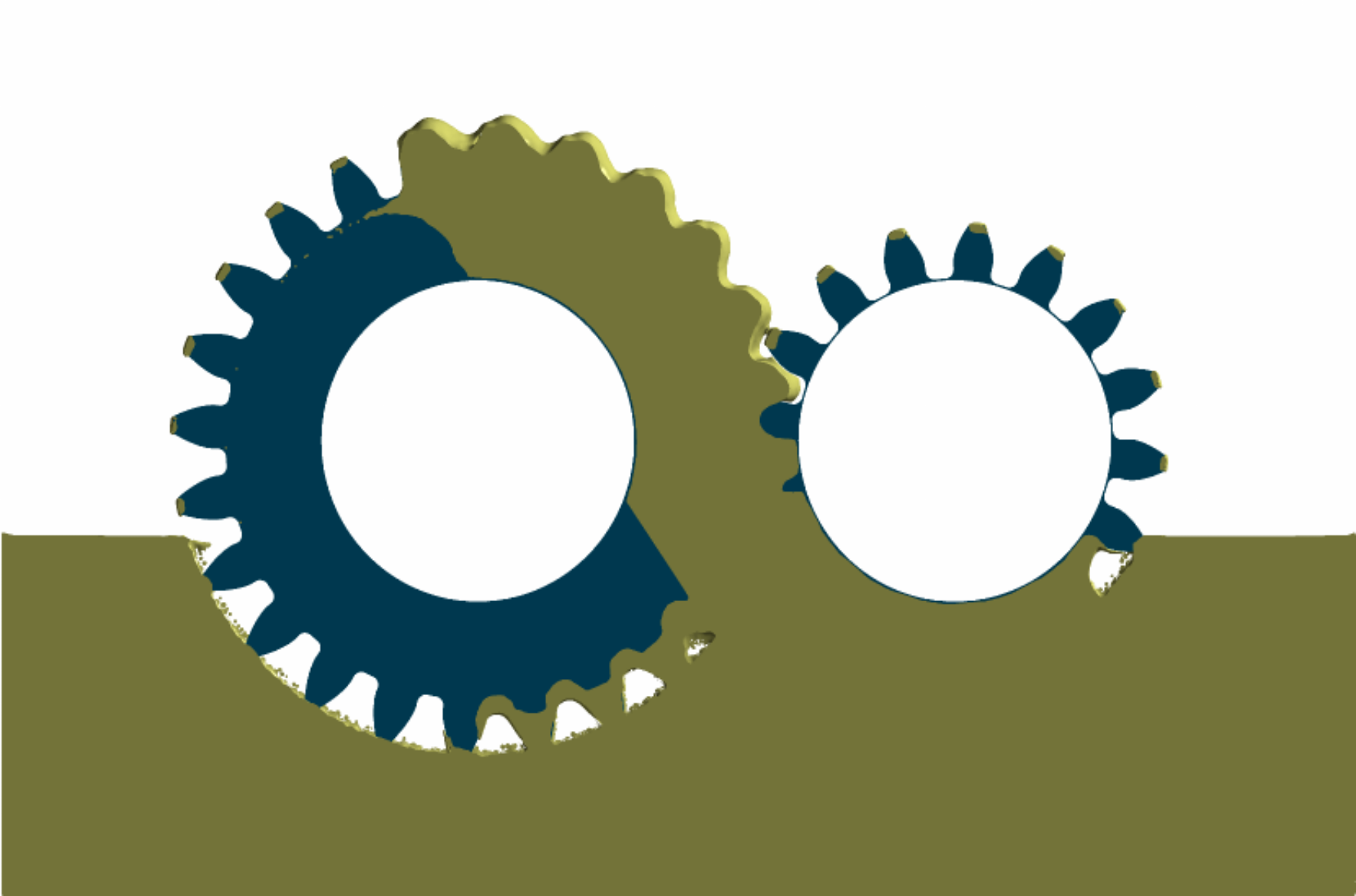}
		\caption{Pinion speed at 100 RPM.}
		\label{fig:FZG_Grease_CaseStudy_100RPM}
	\end{subfigure}
	\label{fig:Mauz_TimeTaken}
        \begin{subfigure}{0.43\textwidth}
		\centering
		\includegraphics[width=\textwidth]{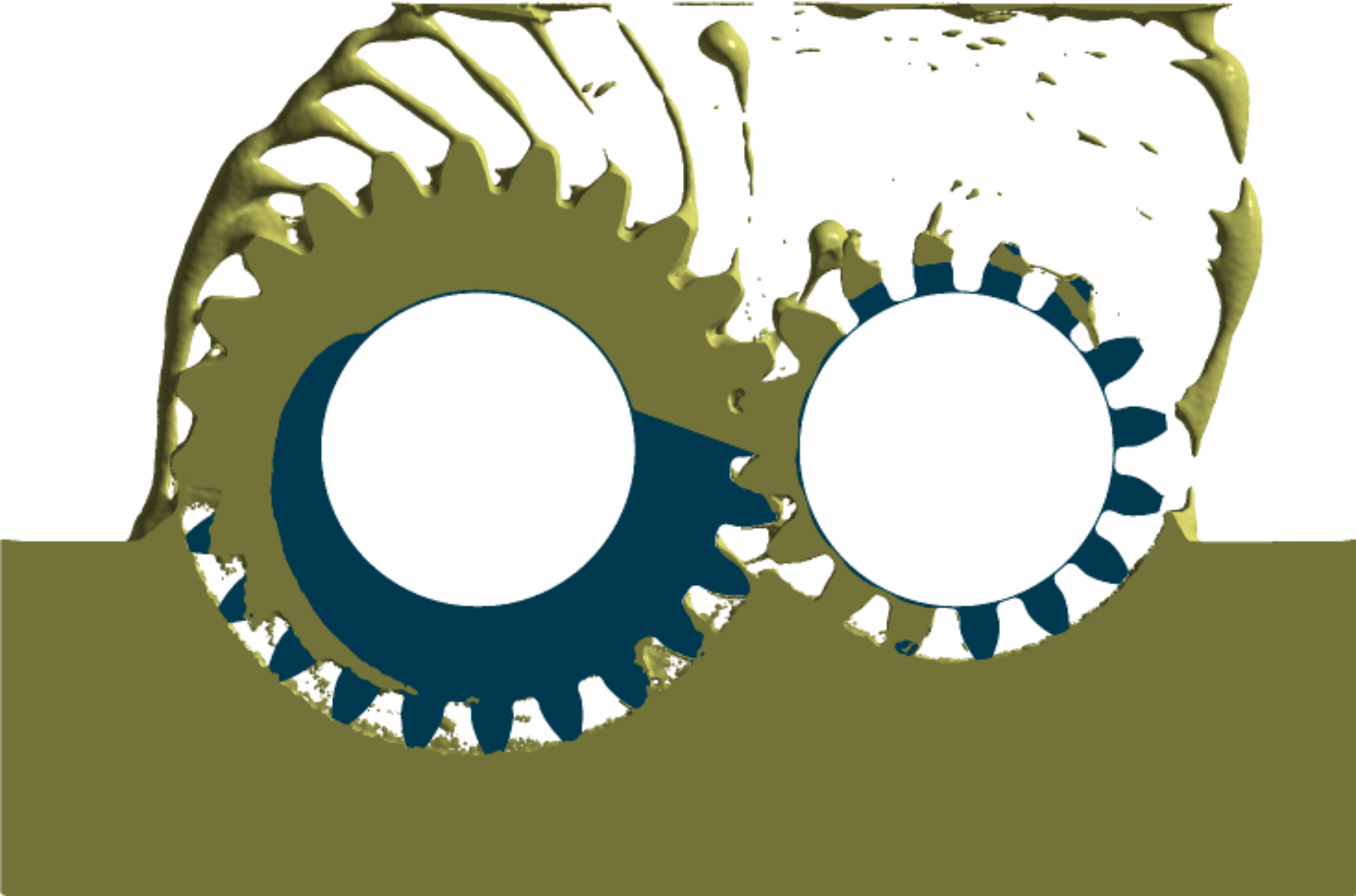}
		\caption{Pinion speed at 1000 RPM.}
		\label{fig:FZG_Grease_CaseStudy_1000RPM}
	\end{subfigure}
	\caption{Top: Quantitative comparison of grease deposition on the pinion and the gear across different rotational speeds. Bottom: Qualitative cut-view visualization of grease distribution, highlighting the influence of rotational speed on deposition patterns.}
	\label{fig:FZG_Grease_CaseStudy_GearSpeed}
\end{figure}
First, the pinion speed is varied from a low value of 13 RPM---the same speed used for validation---up to 5000 RPM. The upper limit is motivated by a wide range of applications such as e-Bikes. For this study, the filling volume is kept constant at 40\%. The amount of grease in the vicinity of gear and pinion is measured after five revolutions for each setting. This simulation results are presented in \cref{fig:FZG_Grease_CaseStudy_GearSpeed}.

Up to a pinion speed of 500 RPM (stage 1), the amount of grease on the gear and pinion remains nearly constant. This behaviour can be attributed to the non-Newtonian and cohesive characteristics of the grease, which promote strong adhesion to the gear and pinion surfaces. At these moderate speeds, the centrifugal forces are insufficient to overcome the internal viscous and adhesive forces of the grease. 

However, as the pinion speed increases beyond 500 RPM, entering stage 2 (500-2000 RPM), the centrifugal forces become strong enough to overcome these resistive forces, resulting in a noticeable reduction in grease volume on the gear and pinion. Specifically, a linear decrease of approximately 40\% in the deposited grease volume is observed across this speed interval, relative to the amount present at or below 500 RPM. By 2000 RPM, the grease volume on the gear decreases by approximately 50\% relative to its value at 1000 RPM. 

At stage 3, speeds above 2000 RPM up to 5000 RPM, a deviation in this linear trend is observed for the pinion. While the grease volume on the gear continues to decrease with a similar linear trend as in stage 2, but the pinion exhibits a slight reduction in the rate of grease loss.  Beyond this point, a near-linear decline in grease volume continues for both gear and pinion surfaces, extending up to 5000 RPM. However, both lose further 50\% of the adhering grease from 2000 to 5000 RPM.

As shown in \cref{fig:FZG_Grease_CaseStudy_1000RPM}, at higher speeds, grease is expelled from the gear and pinion surfaces and adheres to the casing walls, where it remains for the duration of the simulation. The higher the gear speed, the greater the quantity of grease that escapes from the meshing region---the area where the gear and pinion interlock---and is deposited on the housing walls. Furthermore, at all speeds, once the grease is removed from the reservoir, a void forms at that location, as seen in \cref{fig:FZG_Grease_CaseStudy_100RPM}. Due to the limited flowability of the grease, this void remains unfilled over time.

Following the speed study, two speeds---100 RPM and 1000 RPM---are selected to examine the influence of grease filling volume, due to its impact on churning losses. The filling level is varied from 30\% to 80\% of the housing volume. As in the previous analysis, the amount of grease deposited on the gear and pinion is measured after five revolutions. 

\begin{figure}[htbp!]
	\centering
	\begin{subfigure}{0.99\textwidth}
		\centering
		\includegraphics[width=\textwidth]{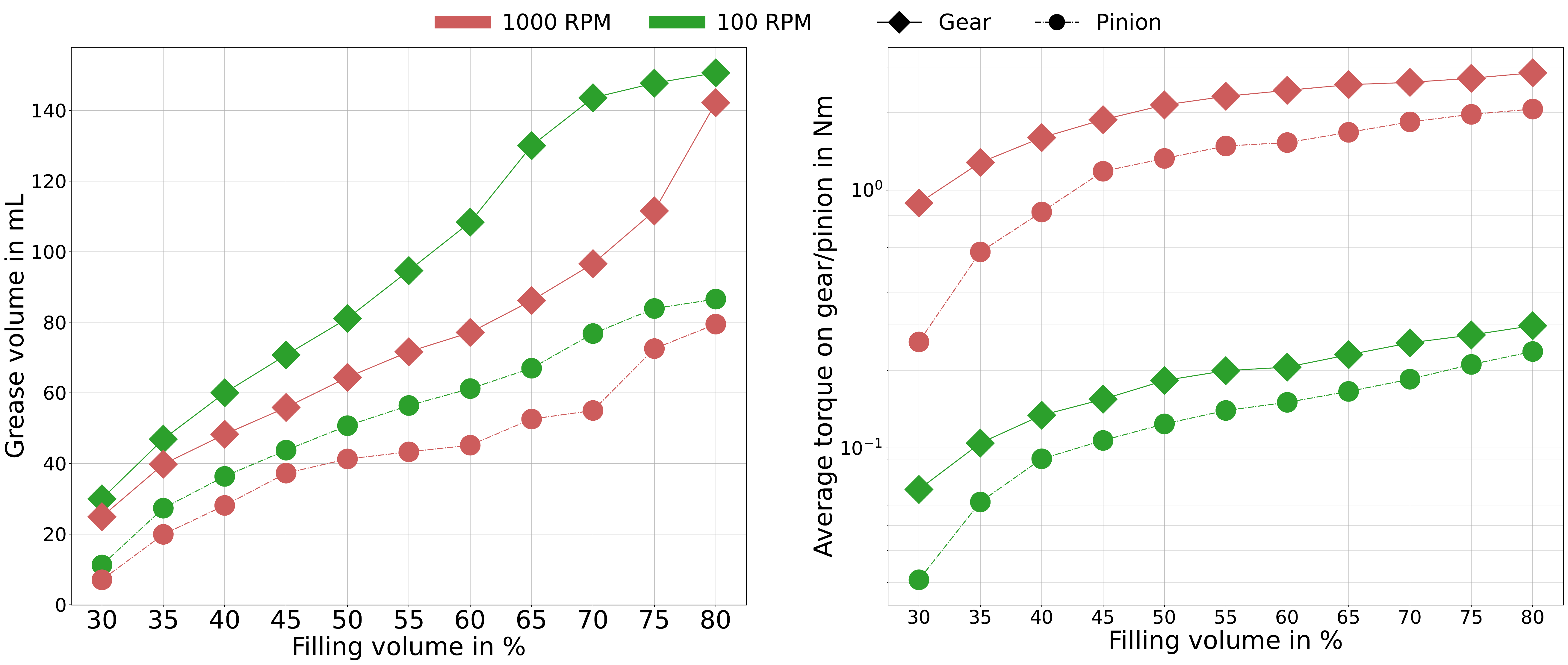}
		\label{fig:PreonLab_GreaseDistribution_Plot}
	\end{subfigure}
        \\
	\begin{subfigure}{0.43\textwidth}
		\centering
		\includegraphics[width=\textwidth]{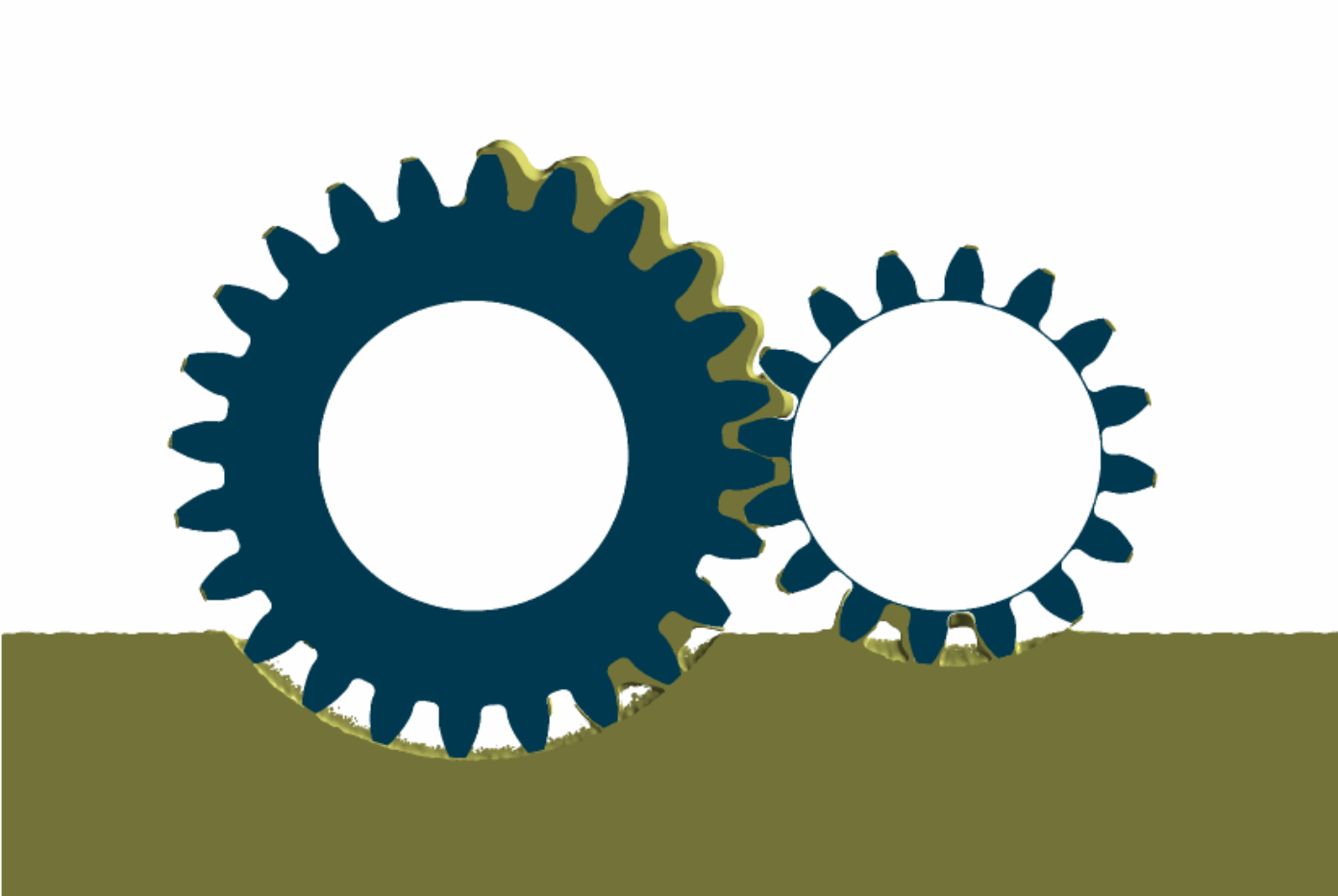}
		\caption{Filling volume = 30\%.}
		\label{fig:FZG_Grease_CaseStudy_100RPM_30Percent}
	\end{subfigure}
        \begin{subfigure}{0.43\textwidth}
		\centering
		\includegraphics[width=\textwidth]{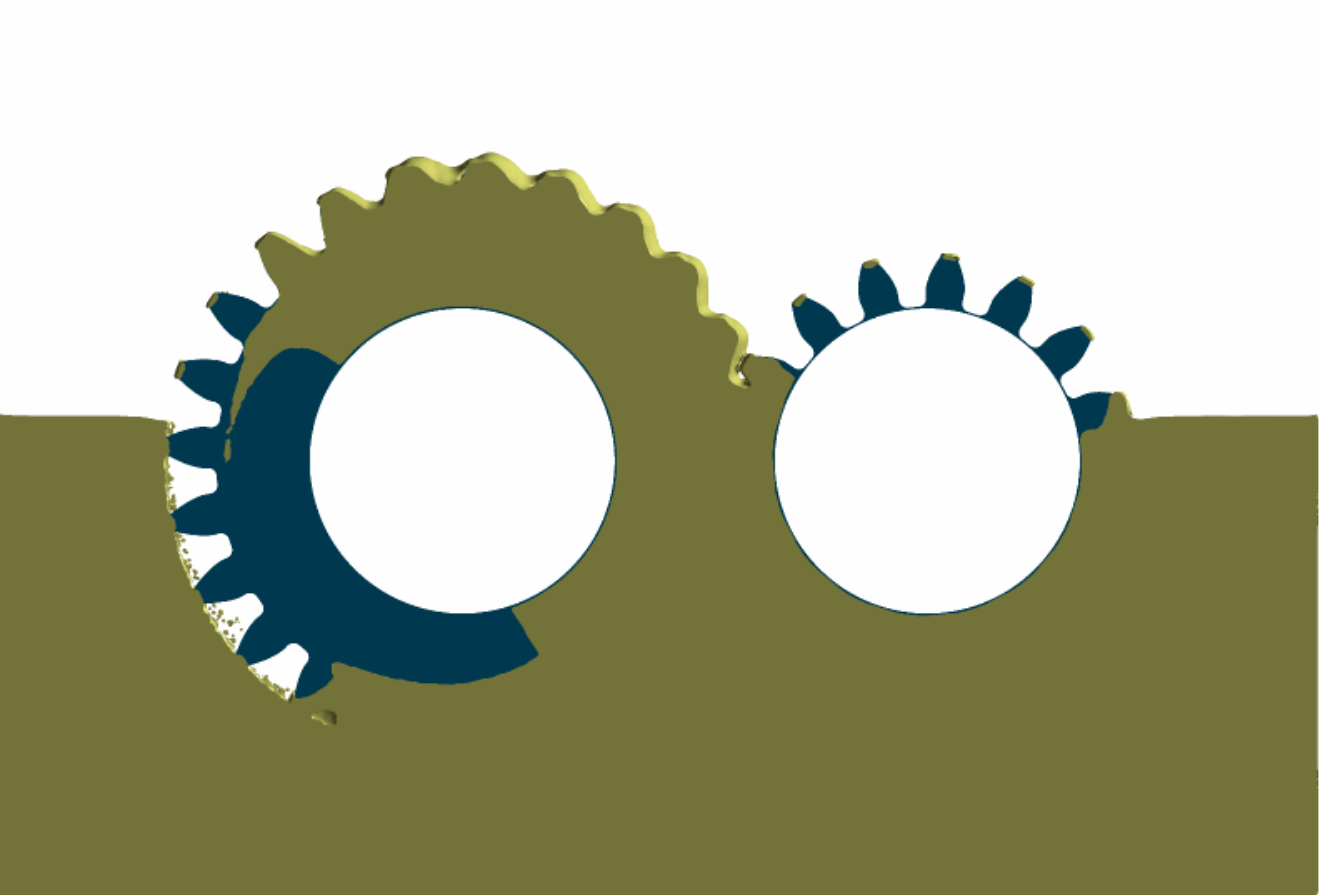}
		\caption{Filling volume = 55\%.}
		\label{fig:FZG_Grease_CaseStudy_100RPM_55Percent}
	\end{subfigure}
        \\
        \begin{subfigure}{0.43\textwidth}
		\centering
		\includegraphics[width=\textwidth]{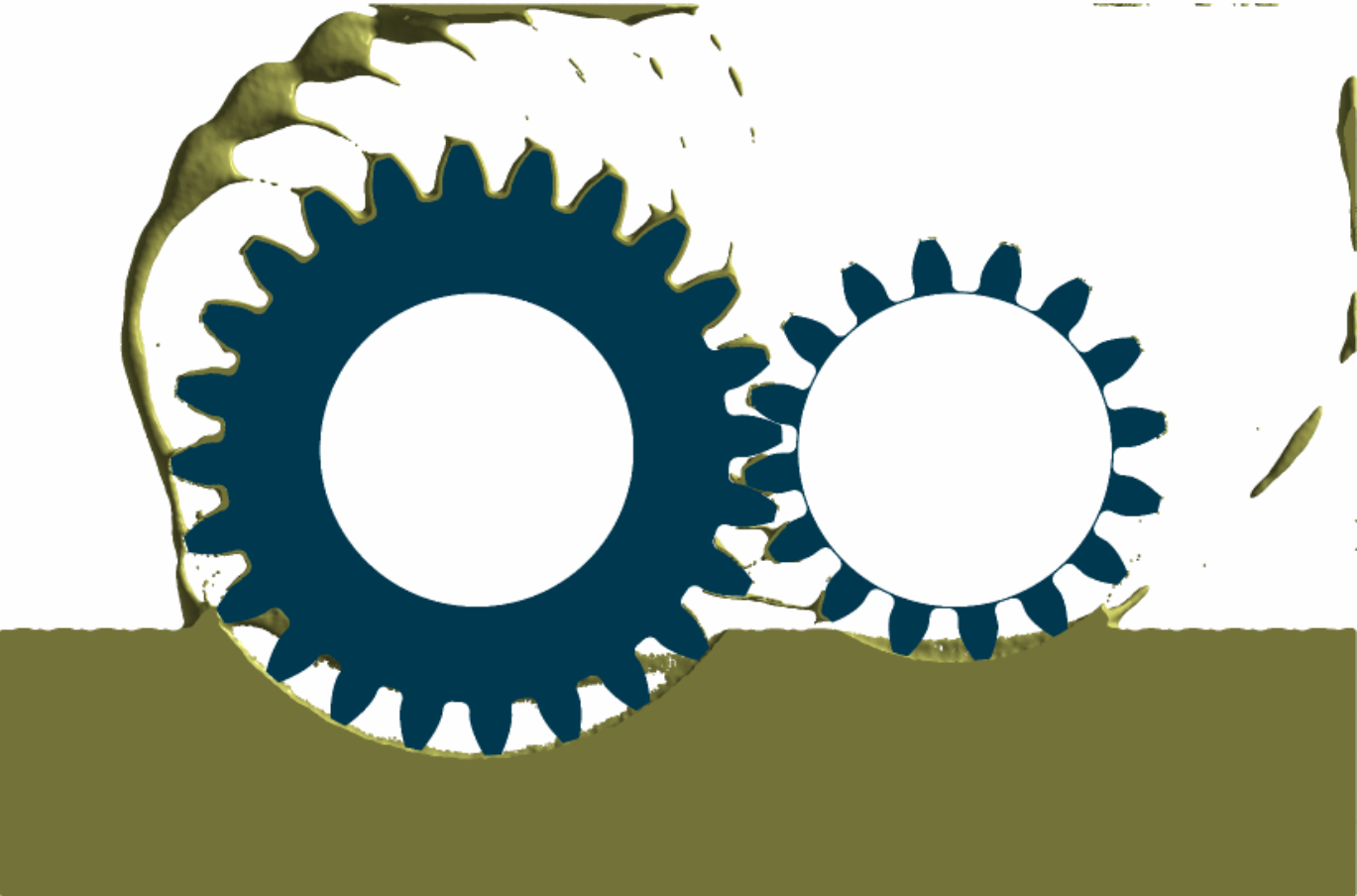}
		\caption{Filling volume = 30\%}
		\label{fig:FZG_Grease_CaseStudy_1000RPM_30Percent}
	\end{subfigure}
        \begin{subfigure}{0.43\textwidth}
		\centering
		\includegraphics[width=\textwidth]{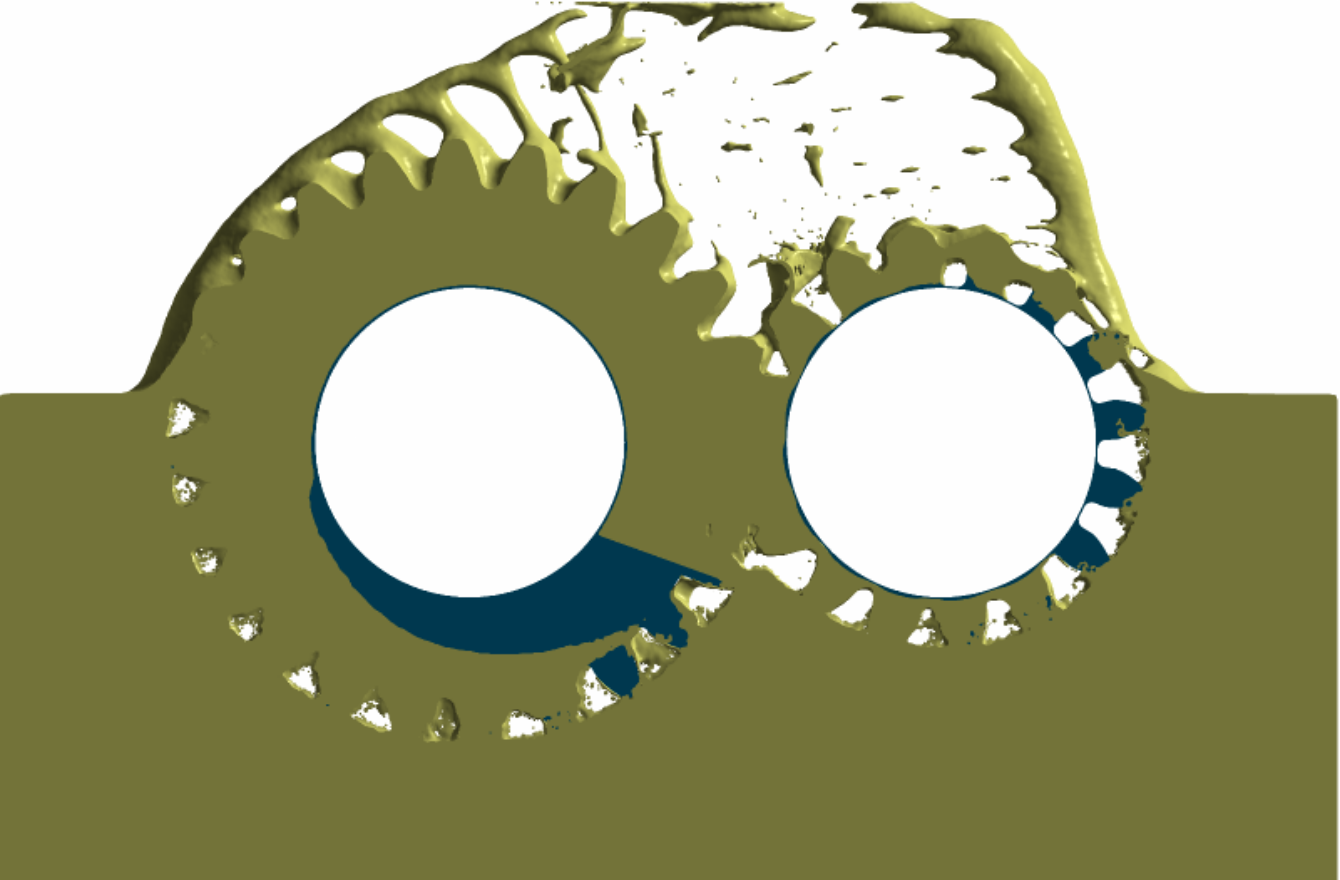}
		\caption{Filling volume = 55\%}
		\label{fig:FZG_Grease_CaseStudy_1000RPM_55Percent}
	\end{subfigure}
	\caption{Top: Quantitative comparison of grease deposition on the pinion and the gear across different filling volumes. Middle: Qualitative cut-view visualization of grease distribution after one revolution at 100 RPM. Bottom: Qualitative cut-view visualization of grease distribution after one revolution at 1000 RPM.}
	\label{fig:PreonLab_GreaseDistribution}
\end{figure}

At both 100 RPM and 1000 RPM, the quantity of grease retained on the gear and pinion increases with increasing filling volume. At lower speeds of 100 RPM, centrifugal forces are small, and the grease remains adhered to the gear and pinion surfaces throughout the simulation, irrespective of the fill level. However, at 1000 RPM, as shown in \cref{fig:FZG_Grease_CaseStudy_1000RPM_30Percent,fig:FZG_Grease_CaseStudy_1000RPM_55Percent}, a low filling level of 30\% results in only a small portion of grease adhering to the gear and pinion, while the majority is flung onto the casing walls, where it remains for the entire duration of the simulation. At an intermediate fill level of 55\%, more grease is retained on the gear and pinion due to stronger adhesive and viscous forces, although some portion is still splashed onto the housing. At the highest fill level of 80\%, the increase in the volume of grease and the associated viscosity resistance help to retain most of the grease in the vicinity of the rotating components. Only a small amount is splashed onto the housing. 

In addition to the amount of grease deposited, the plots in \cref{fig:FZG_Grease_CaseStudy_GearSpeed,fig:PreonLab_GreaseDistribution} also show the simulated average torque acting on the pinion and the gear. It is important to note that the torque values shown in this study represent the contributions from the grease phase only; the influence of the air phase is not accounted for by the solver. The torque acting on the gear and pinion increases with rotating speed and filling volume. Despite a reduction in grease coverage at higher speeds, the initial and average torque tend to increase with speed. Increasing fill levels enhance lubrication coverage, promoting more effective wetting of the gear and pinion surfaces. However, this also results in greater churning losses and increased torque resistance. To accurately quantify and validate the simulated torque losses, further experimental studies with direct torque measurements for grease lubricated gearboxes are essential.

\newpage
\section{Summary and Conclusion}
\label{sec:conclusion}

This study presents a numerical investigation of lubricant distribution within gearboxes, comparing four distinct solvers: the mesh based OpenFOAM and Ansys-Fluent, and the particle based PreonLab and MESHFREE. A combination of qualitative and quantitative validations against experimental data is used to assess solver accuracy and computational efficiency.

For qualitative validation, the simulated lubricant distribution patterns are compared with the experimental results. All solvers show good agreement, with PreonLab providing the closest match while also achieving the shortest simulation times. This is followed by mesh-based solvers, which generally require significantly longer runtimes.

For quantitative validation, the torque acting on a single gear immersed in an oil sump is compared between simulations and experiments. OpenFOAM produces the most accurate results for low-viscosity lubricants, but show greater deviation for high-viscosity cases. Ansys-Fluent maintains consistent performance, with deviations ranging between 1\% and 15\%. While PreonLab produces a higher deviation (17\%) for low-viscosity lubricants, it outperforms the other solvers for high-viscosity lubricants, with deviations as low as 3\%. MESHFREE exhibits approximately 50\% deviation in all cases. In terms of computational efficiency, PreonLab is the most efficient, followed by MESHFREE, OpenFOAM, and Ansys-Fluent.

Given that many industrial gearboxes are lubricated with grease, PreonLab is further employed to investigate grease distribution. The solver is first qualitatively validated against experimental observations, showing good agreement, although the adhesion and cohesion parameters must be adjusted for each grease. The influence of gear speed and filling volume on the amount of grease deposited on the gears is then analysed. The results shows that up to a certain gear speed threshold, the grease amount on the gears remains nearly constant; beyond this threshold, the deposited grease volume decreases with increasing gear speed. Additionally, for two selected gear speeds, the amount of grease on the gears increases with higher filling volumes.

The results obtained in this work highlight the strengths and limitations of each solver, providing practical guidance for selecting the appropriate tools to predict lubricant flow and torque in gear systems. In addition, the validation of PreonLab for grease distribution demonstrates its capability to simulate such flow in complex rotating systems. For gearboxes using grease, this study also identifies a threshold gear speed below which the grease distribution remains stable. Furthermore, this workflow with PreonLab offers potential for further investigations into grease behaviour under varying rheological properties and operational conditions, including the identification of critical speed thresholds across different grease types.

\appendix

\label{sec:sample:appendix}

\bibliographystyle{elsarticle-num} 
\bibliography{cas-refs}





\end{document}